\documentclass[useAMS,usenatbib]{mnras}
\usepackage{graphicx}
\graphicspath{{images_de_larticle/}}
\usepackage{amsmath}

\usepackage[usenames,dvipsnames]{color}

\def\lya{Ly$\alpha$~}

\title[Large opacity fluctuations in the \lya forest  at $z \sim 5.5-6  $]{
Large scale opacity fluctuations in the \lya forest:  evidence for QSOs dominating the ionizing UV background at $z\sim 5.5-6$? 
}
\author[J. Chardin et al.]{Jonathan Chardin$^{1}$\thanks{E-mail:jc@ast.cam.ac.uk}, 
Ewald Puchwein$^{1}$ and Martin G. Haehnelt $^{1}$ \\
$^{1}$Kavli Institute for Cosmology and Institute of Astronomy, University of Cambridge, Madingley Road, Cambridge, CB3 0HA, UK\\}

\begin{document}


\date{Accepted / Received }

\pagerange{\pageref{firstpage}--\pageref{lastpage}} \pubyear{2016}

\maketitle

\begin{abstract}
Lyman-alpha forest data probing the post-reionization Universe shows surprisingly  large opacity fluctuations over rather large ($\ge 50 h^{-1}$ comoving Mpc) spatial scales.   
We model these fluctuations using a hybrid approach utilizing the large volume Millennium simulation to predict the spatial distribution of QSOs combined with  smaller scale full hydrodynamical 
simulation performed with RAMSES and post-processed  with the radiative transfer code ATON. We  produce realistic mock absorption spectra that account for the contribution 
of galaxies and QSOs  to the ionizing UV background.   These improved models confirm our earlier findings  that a significant ($\ga$50\%) contribution of ionizing photons from QSOs can  
explain the  large reported opacity fluctuations on large scales. The inferred QSO luminosity function is thereby  consistent with recent estimates  of the space 
density of QSOs at this redshift. Our simulations still  somewhat struggle, however,   to reproduce the  very  long $(110  h^{-1}$ comoving Mpc)  high opacity absorption  
through observed in ULAS J0148+0600, perhaps suggesting an even later end of reionization than assumed in our previously 
favoured model. Medium-deep/medium area QSO surveys  as well as targeted searches for the predicted strong transverse QSO proximity effect would   
illuminate the origin of the observed large scale opacity fluctuations. They would allow to substantiate whether UV fluctuations due to QSO are indeed primarily responsible, 
or whether significant contributions from other recently proposed mechanisms such as large scale fluctuations in temperature and mean free path (even in the absence of rare bright sources) are required. 
\end{abstract}



\section{Introduction}
\label{intro}

The \lya forest is  the primary  probe of the ionization state  of hydrogen in the post-reionization Universe (see \citealt{2015PASA...32...45B} for a recent review). 
The  observed rapid rise  in the  mean flux level in \lya forest data at  $z <6 $ almost certainly marks  the end of the reionization epoch
(\citealt{2006AJ....132..117F}; \citealt{2013MNRAS.430.2067B}; \citealt{2015MNRAS.447.3402B}), 
while  the rapid decline of the occurrence of \lya emitters at $z>7$ (\citealt{2013MNRAS.429.1695B} and \citealt{2014MNRAS.440.3309D}) 
and the rather low value of the Thomson optical depth suggested by recent Planck data \citep{2015arXiv150201589P,2016arXiv160503507P}
point to  a more rapid progress and a later start of the reionization of hydrogen than  suggested by the early WMAP results.  
The \lya forest data thereby does not only  constrain  when reionization ended,  but may also provide  
important clues to the nature of the ionizing sources (see \citealt{2001MNRAS.321..450T}, \citealt{2007ApJ...670...39L}, 
\citealt{2008MNRAS.383..691W}, \citealt{2013ApJ...775...81R}, \citealt{2015ApJ...812...30K} 
and reviews of \citealt{2005SSRv..116..625C}, \citealt{2009RvMP...81.1405M}, \citealt{2015PASA...32...45B} and \citealt{2015arXiv151200086M}). 
  
The two types of ionizing sources considered to be most relevant during the reionization epoch are 
massive stars in star-forming galaxies and accreting supermassive black holes, while emission from accreting 
compact objects in  binary stars is normally considered to be less important. More exotic sources, like decaying dark matter particles or evaporating primordial black holes have been also suggested to  contribute to the ionizing photon budget. 
In recent years most works in the field favour star-forming galaxies as the main driver of reionization.
Their luminosity function is now well measured up to z$\sim$10 (see \citealt{2015ApJ...803...34B}) down to a magnitude of $\mathrm{M_{AB \, 1600}\sim-16}$.
However, the fraction of radiation that actually escapes from star-forming galaxies into the intergalactic medium $\mathrm{f_{esc}}$ is still poorly constrained.
In the local Universe only a few starburst galaxies are observed as Lyman continuum leakers with $\mathrm{f_{esc}\le5\%}$  
(see \citealt{2014Sci...346..216B} and \citealt{2016ApJ...823...64L}).
Efficient ionizing radiation leakers ($\mathrm{f_{esc}\ge10\%}$) are more numerous at higher redshift (\citealt{2016arXiv160603452R}),  but only a few are robust detections 
(\citealt{2016ApJ...826L..24S} and  \citealt{2016arXiv160605309L}). 
Overall, even at high redshift, the average escape fraction in all observations is still very small with $\mathrm{f_{esc}\le2\%}$ (\citealt{2015ApJ...804...17S}, 
\citealt{2015ApJ...810..107M} and \citealt{2016A&A...585A..48G}).
Recent numerical studies suggest that, on average, only $\sim$ 10\%
of the Lyman continuum photons escape from their host haloes in the mass range
$\mathrm{10^8M_{\odot}<M_{halo}<10^{11}M_{\odot}}$ (\citealt{2014ApJ...788..121K}, \citealt{2015MNRAS.453..960M} and \citealt{2016arXiv160407842X}). 
Only mini-haloes might reach high average escape fractions of $\mathrm{f_{esc}\sim20-40\%}$ (\citealt{2014MNRAS.442.2560W}, \citealt{2016arXiv160407842X} and 
\citealt{2016arXiv160804762K}), but it has been claimed that they would be a minor contributor to reionization as their star formation is inefficient (\citealt{2016arXiv160804762K}).
Therefore, even for an optimistic assumption of $\mathrm{f_{esc}\sim20\%}$, the luminosity function of star forming galaxies has to be extrapolated to
faint magnitudes $\mathrm{M_{UV}\sim -13}$ to provide enough ionizing photons to completely 
reionize the Universe by z=6 (\citealt{2013ApJ...768...71R} and \citealt{2015ApJ...802L..19R}).

On the other hand it has been claimed  many times that emission from QSOs  alone is  likely not sufficient  to reionize the Universe completely by z=6  (see e.g. the pioneering work of \citealt{1999ApJ...514..648M}).
In the last update of their model \citealt{2012ApJ...746..125H} argued once more  that quasars alone would fail to reionize the Universe.
The observations used to constrain the UV luminosity function of QSOs during reionization (\citealt{2007ApJ...654..731H}) suggested  a ionizing  emissivity of QSOs   well below the one required for reionization by z=6 even assuming $\mathrm{f_{esc}=1}$.
However, the observed QSO  luminosity function at $z>4$ is still rather uncertain, in particular  around the "knee" of the luminosity function 
that should dominate the total ionizing emissivity from QSOs. 
The luminosity function used in calculations of the ionizing emissivity has therefore often been extrapolated from results at lower redshift.
Recently, \citealt{2015AA...578A..83G}  used a  X-ray selection technique to provide a new observational estimate   of the space density of   faint QSOs in  the CANDELS GOODS-S survey just at the tail end of reionization (z$\sim5.75$).
Surprisingly they found that their estimated space density was significantly higher 
than most extrapolations of the QSO luminosity function from lower redshift.
This led \citealt{2015ApJ...813L...8M} to reconsider the ionizing photon budget from QSOs and they found that, if the increased ionizing 
emissivity suggested by    \citealt{2015AA...578A..83G}  can be extrapolated to higher redshift,  QSOs  could  potentially reionize the Universe alone, even without a contribution from star-forming galaxies.

In a recent paper \citet{2015MNRAS.447.3402B} (B2015 hereafter) presented detailed measurements of the 
probability distribution function (PDF) of the \lya effective optical depth $\mathrm{\tau_{eff}}$ averaged over scales 
of 50 cMpc/h in the redshift range $4\le z \le 6$ based on a sample of high S/N high-resolution  QSO absorption spectra. 
B2015 found rather large fluctuations of the mean flux  at $z>5.5$.
There are now five theoretical scenarios that have been suggested to  explain the large scatter in the opacity distribution at the tail-end of reionization.
First B2015, argued that opacity fluctuations would result from fluctuations in the mean free path of ionizing photons 
just after percolation of HII regions.
Second, right after B2015, \citet{2015MNRAS.453.2943C} pointed out 
that fluctuations in the UV background from bright sources such as faint AGNs could cause the right amount of opacity fluctuations.
Third, \citet{2015ApJ...813L..38D} invoked large fluctuations of the temperature due to patchy reionization that could also lead to the observed opacity fluctuations.
Fourth, \citet{2016MNRAS.460.1328D} claimed that large scale variations in the spatial distribution of the mean free path of ionizing photons in a galaxy only scenario 
would generate the right amount of fluctuations.
Fifth, recently, \citet{2016arXiv160503183G} argued that their set of radiation hydrodynamic simulations of reionization naturally gives the observed opacity fluctuations 
when averaging multiple simulations with different initial conditions corresponding to different mean densities on the simulation box scale.

In the latter case, different reionization histories due to the different mean densities naturally lead to a level of scatter from line-of-sight to line-of-sight that matches the observed scatter.
However, it is not clear that averaging from multiple simulations with different reionization histories is not introducing a bias in the results as the different simulated boxes evolve independently,
while in reality underdense regions of the Universe may be reionized by neighbouring denser regions.
Therefore the results would likely differ from a single simulation with the same effective volume.
The spatial variation of the mean free path case (B2015 and \citealt{2016MNRAS.460.1328D}) is in tension with the scenario of \citet{2015ApJ...813L..38D}.
Indeed, large values of the mean free path would be expected near overdensities containing many sources and much smaller values in voids far away from them.
This would lead to high transmissivity peaks in the \lya forest near overdensities and to opaque segments in voids in these models.
Conversely, in the temperature fluctuation model of \citet{2015ApJ...813L..38D}, we would expect colder regions near overdensities as those places are the first to reionize and have a longer time to cool after a passing ionization front.
Thus, high transmissivity peaks are expected far away from the galaxies, in voids that have been reionized at the end of the process.
Finally, our model of UV background fluctuations due to rare bright sources would lie in the middle of the two latter scenarios.
High transmissivity regions are expected close to bright sources as in the fluctuating mean free path model but low transmissivity 
regions could also occur near galaxy location that are far away from bright sources as in the temperature fluctuation model.
Therefore, all the existing theoretical models lead to different observational signatures and forthcoming surveys could potentially discriminate between them.
In the meantime, a more robust modelling is also required to provide guidance in interpreting the observational results.

Modelling the large scale topology  of reionization  realistically is, however,  
very difficult due to the large dynamic range required (see \citealt{2014ApJ...793...29G}, \citealt{2014ApJ...793...30G} and \citealt{2015arXiv151100011O}  
for recent  ambitious attempts) and the numerically expensive coupling of radiative transfer effects and gas dynamics that is increasingly 
added to the simulations codes (see \citealt{2013MNRAS.436.2188R}, \citealt{2015MNRAS.451.1586P}, \citealt{2015MNRAS.452.2034R} and \citealt{2015MNRAS.454.1012A}).
Recently we performed radiative transfer simulations of the reionization of 
hydrogen by (faint) galaxies in post-processing. They (marginally) resolve the sinks of ionizing radiation and show only rather moderate fluctuations of the UV background and the mean free 
path in the post-overlap phase of reionization (\citealt{2015MNRAS.453.2943C}).
As already said, this led \citet{2015MNRAS.453.2943C} to suggest that much rarer brighter sources  like QSOs  with space densities of 
$\mathrm{\sim 10^{-6} (cMpc/h)^{-3}}$ may contribute significantly to the ionizing background at $z\sim5.5-6$ and be primarily responsible for the substantial opacity 
fluctuations at scales of 50 cMpc/h and beyond at this redshift. 
We investigate here in detail the implications of  this possible explanation of the large opacity fluctuations on large 
scales for the QSO luminosity function and the  contribution of QSOs to the ionizing 
UV background at $z>5$ (\citealt{2015AA...578A..83G}; \citealt{2015A&A...575L..16H}; 
\citealt{2015ApJ...813L...8M}, \citealt{2015MNRAS.453.1946G}, \citealt{2016arXiv160204407Y}, \citealt{2016arXiv160302281M} \citealt{2016arXiv160303240F}).

This paper is organized as follows. In Sect. \ref{Simu} we present our hybrid approach utilizing the large volume Millennium 
simulation to model the spatial distribution of QSOs 
combined with smaller 
scale full hydrodynamical simulations performed with RAMSES and post-processed  
with the radiative transfer code ATON. 
Sect. \ref{results} presents our results with regard to spatial fluctuations of the photoionization rate and the corresponding opacity fluctuations in the \lya forest.  
Sect. \ref{discussion} discusses our work in the context of alternative models  to explain the opacity fluctuations observed on large scales. 
We  give our conclusions and outlook in Sect. \ref{prospects}.

\begin{figure}
  \begin{center}
     \includegraphics[width=9cm,height=7cm]{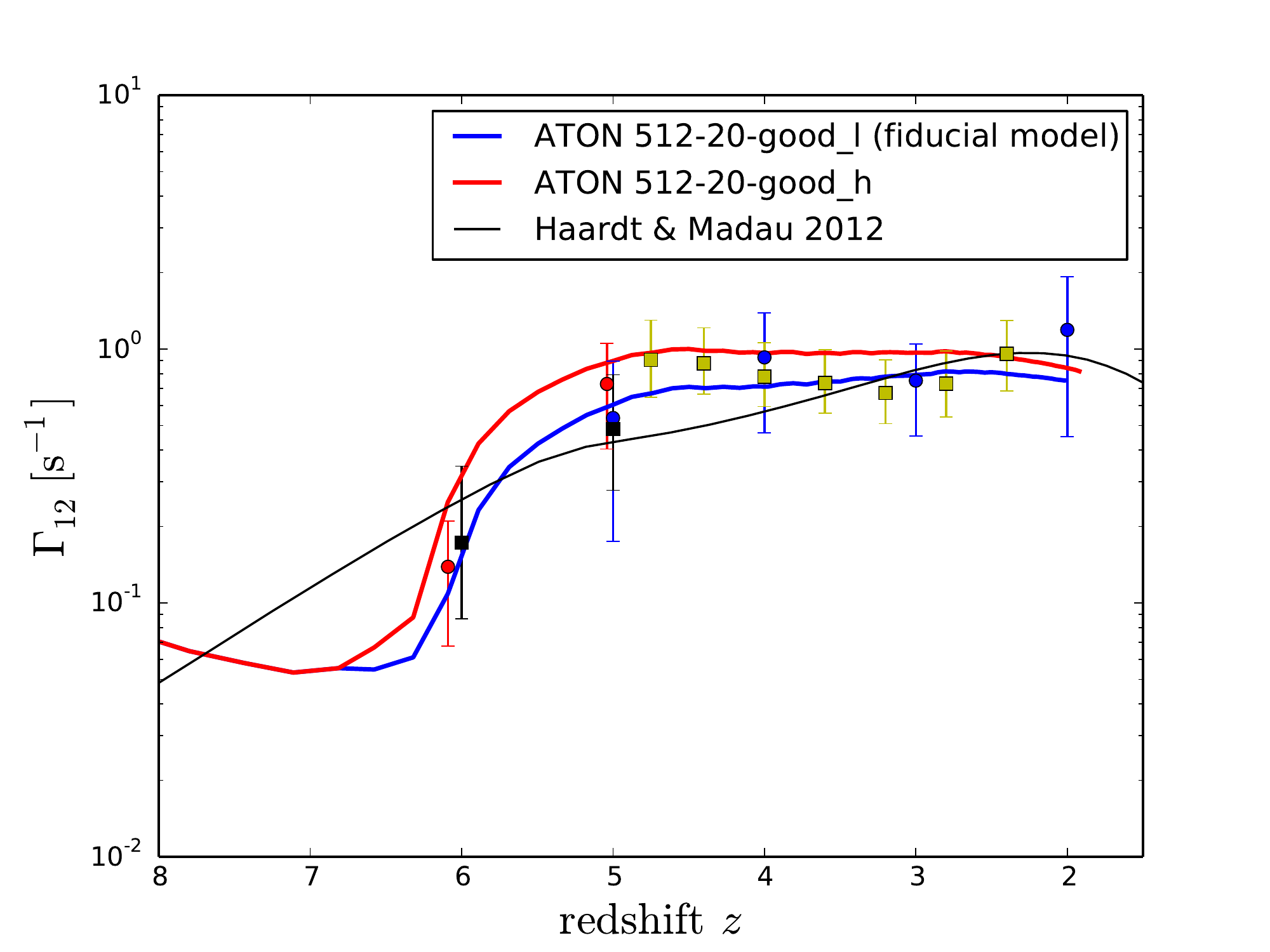} 
  \caption{Redshift evolution of the average photoionization rate in already ionized regions 
(with an ionization fraction $x \ge 0.5$) in the full  radiative transfer simulations from paper I.
The observational constraints shown with the different colored symbols with errorbars are from \citet{2007MNRAS.382..325B} (blue), 
\citet{2011MNRAS.412.2543C} (red), \citet{2011MNRAS.412.1926W} (black) and \citet{2013MNRAS.436.1023B} (yellow).}
    \label{gamma_two_models}
  \end{center}
 \end{figure}

\section{Methodology}
\label{Simu}

In this section we will describe our  modelling  of the impact of large-scale fluctuations of the photoionization 
rate $\Gamma$ on the PDF of the high-redshift \lya optical depth. First we will present briefly the  main properties of the full
(post-processed) radiative transfer hydrodynamical simulations  used in our recent study (\citealt{2015MNRAS.453.2943C}, hereafter paper I). 
Then we will describe  our  model  of  the spatial distribution of the contribution of QSO  to the ionizing background based on a hybrid approach 
utilizing both the large volume Millennium dark matter simulation  (\citealt{2005Natur.435..629S}) as
well as our smaller volume full (post-processed) radiative  transfer hydrodynamical simulations.

\subsection{The radiative transfer simulations} 
\label{simulation}

\subsubsection{The cosmological hydro-simulations} 
\label{Hydro}

The simulations from paper I are performed in two steps.
First hydrodynamic simulations are run. Then in the second step the radiative transfer calculations are performed 
in post-processing on top of the hydrodynamic simulations adopting an ionizing source model.

The (cosmological) simulations of the evolution of the dark matter and the hydrodynamics of the gas 
were performed  with the RAMSES code (\citealt{2002A&A...385..337T}).
Note that we did not employ  the adaptive mesh refinement (AMR) option of  RAMSES. 
We  utilized  a coarse, fixed  grid discretized in $512^3$ cells. The  uniform UV background model of \citet{2012ApJ...746..125H} 
(HM2012 hereafter) was implemented in RAMSES to account for the photoheating of the gas.
This gives a temperature evolution of the gas at mean density in reasonable agreement with recent observations of \citealt{2011MNRAS.410.1096B} and 
\citealt{2014MNRAS.441.1916B} (see Fig. 3 of paper I).

Note that our deliberately  simplified simulations of the IGM also did not take into account supernovae feedback and metal enrichment/cooling.
As noted in paper I, the lack of supernovae feedback mainly affects the neutral hydrogen distribution 
in galactic haloes which are too spatially concentrated to reproduce the observed incidence rate of dense absorbers.

\subsubsection{Post-processed radiative transfer simulations with ATON} 
\label{postRT}

The radiative transfer calculations were performed in  post-processing with the ATON code (\citealt{2008MNRAS.387..295A}).
ATON is a GPU based radiative transfer code utilizing a moment based description of the radiative transfer equation.
We employed  a monochromatic treatment that assumes all ionizing photons have an energy of 20.27 eV.
The choice of energy of 20.27 eV is described in \citet{2009A&A...495..389B}. 20.27 eV is the mean energy of a 50000 K blackbody 
spectrum that represents well the emission of ionizing photons of  a population of stars with a Salpeter initial mass function in the mass range $1 M_\odot-100 M_\odot$.
Adopting a different energy would only mildly affect  the  spatial distribution of the ionization fraction.
A different photon energy  will mainly  affect the temperature calculations, as photons with a higher energy would heat the gas more efficiently. 
However, the temperature calculation in our radiative transfer module is not used in  our study.  The temperatures are instead  taken from the hydro-simulation 
performed with a HM2012 UV background accounting for the ionizing spectrum of galaxies and quasars. 
This results in a reasonable thermal history of the IGM compared to observations. 
Full radiative transfer ATON simulations  were created  based on snapshots of the  optically thin RAMSES  simulations separated by 40 Myrs. 

Ionizing sources were placed in  the dark matter haloes identified in the RAMSES simulation
and assumed to emit continuously. The ionizing luminosities were  calibrated similarly as in  \citet{2006MNRAS.369.1625I} 
(but see also \citealt{2012A&A...548A...9C} and \citealt{2014A&A...568A..52C}) assuming a linear scaling 
of the ionizing luminosity with the mass of dark matter haloes. The normalisation is thereby assumed to  
vary with redshift and is chosen  so that  the integrated  comoving ionizing emissivity is roughly similar to that of the HM2012 
uniform UV background model. Slight modifications were allowed so as to obtain an improved match of the 
hydrogen photoionization rates inferred from \lya forest data.

In figure \ref{gamma_two_models} we show $\Gamma_{12}(z) \equiv \Gamma(z) / 10^{-12}$  for  the `512-20-good\_l' and the `512-20-good\_h'
models from paper I that both agree well with  the measured photoionization rates in the post-reionization Universe.
In the acronyms, 512 denotes  the  resolution of $512^3$ for the (coarse) grid while 20 stands for the box size of 20 comoving Mpc/h. 
In the particular case of these two simulations, ``\_l'' and ``\_h'' denote respectively ``low'' and ``high'', where low means 
a lower value of $\Gamma_{12}$ and high a higher value of $\Gamma_{12}$ at $z\sim6$.
In the rest of the paper we will extensively use the `512-20-good\_l' model in our combined model of galaxies and AGNs, which we introduce in the next section.

\subsection{A combined UVB model for galaxies and AGNs}
\label{galaxy_UVB_model}

\renewcommand{\arraystretch}{1.5}
\setlength{\tabcolsep}{0.5cm}
\begin{table*}
\begin{center}
\begin{tabular}{lc|c|c|c|c|c|}
  \hline
  $\mathrm{log}_{10}(\phi^*)$ & $M_{*}$ & $\beta$ & $\gamma$ & $N_{\rm QSO}$ &$\epsilon_{912} \, [10^{24} \mathrm{erg/s/Hz/Mpc^3}]$  \\
  \hline
 -5.8 & -23.4 & 1.66 & 3.35 & 1296 & 1.31   \\
  \hline 
\end{tabular}
\caption{Parameters for the two-power law fit of the QSO luminosity function of equation \ref{power_law_fit} 
that fits the data of \citet{2015AA...578A..83G}. $N_{\rm QSO}$ is  the number of QSOs above a limiting magnitude of $\mathrm{M_{AB}=-22}$ 
in  the $500^3$ $\mathrm{(Mpc/h)^3}$ volume of the Millennium simulations 
and  $\epsilon_{912}$ is the corresponding comoving ionizing 
emissivity integrated over magnitudes brighter than  $\mathrm{M_{AB}=-22}$.}
\label{tab1}
\end{center}
\end{table*}

\renewcommand{\arraystretch}{1.5}
\setlength{\tabcolsep}{0.5cm}
\begin{table*}
\begin{center}
\begin{tabular}{|c||c|c|c|}
  \hline
  redshift $z$ & 5.8 & 5.6 & 5.4  \\
  \hline
 fixed  $\lambda_{\mathrm{mfp}}^{912}$ [cMpc/h]& 15,25,35 & 33,43,53 & 44,54,64    \\
  \hline
$\lambda_{0}$ [cMpc/h]& $\mathrm{18.0}$ & $\mathrm{21.6}$ & $\mathrm{26.0}$  \\
\hline
$\Gamma_{0}$ [$\mathrm{s^{-1}}$]& $\mathrm{2.95\times10^{-13}}$ & $\mathrm{3.35\times10^{-13}}$ & $\mathrm{3.74\times10^{-13}}$  \\
\hline
\end{tabular}
\caption{The range of $\lambda_{\mathrm{mfp}}^{912}$ values investigated in our model with a constant mean free path for all the three redshift bins. 
The values of $\lambda_{0}$ and $\Gamma_{0}$
are the normalisation used in equation \ref{furlanet_gamma} for our model with a $\Gamma$-dependent mean free path. }
\label{tab2}
\end{center}
\end{table*}

In Paper I we had seen that  while our full radiative models of reionization by (faint) galaxies  were able to reproduce the mean 
flux levels in the observed  spectra, they did not reproduce the rather wide distribution of optical depths as measured by 
B2015 for 50 cMpc/h  chunks at $z>5$. We  thus investigated   a simple toy model in which we assumed that a small 
number of bright sources assumed to be located in the most massive haloes of  a $100^3$ $\mathrm{(Mpc/h)^3}$ DM simulation contribute to the ionizing UV background. 
We found that a model where about half of the ionizing UV background is due to bright ionizing  sources with  a space density  
of about $10^{-6} $ (cMpc/h) $^{-3}$ reproduces the PDF of the flux as measured by B2015 well.   

\begin{figure}
  \begin{center}
     \includegraphics[width=9cm,height=7cm]{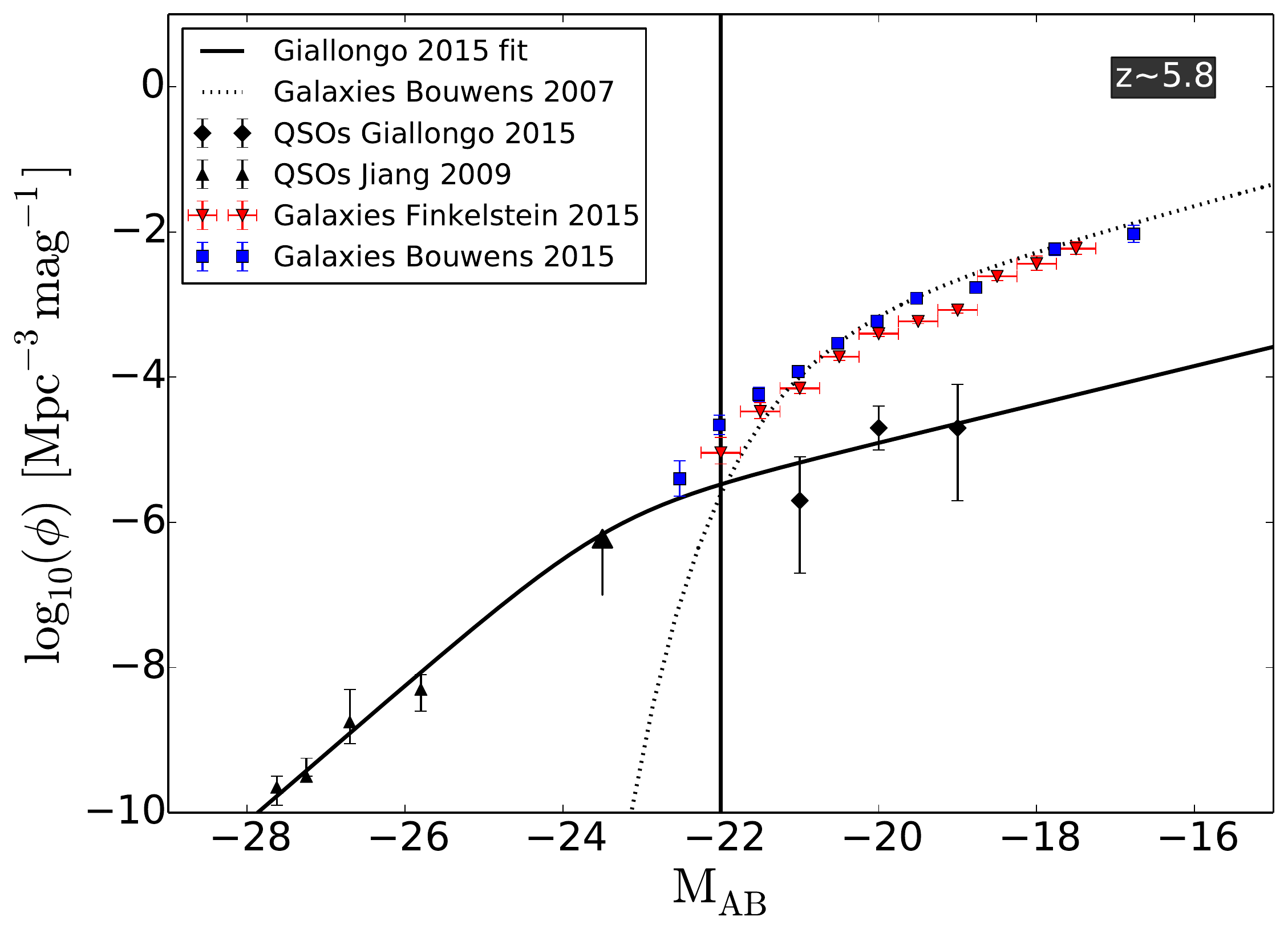} 
  \caption{Fit to the QSO luminosity function (solid black) obtained  by  \citealt{2015AA...578A..83G}
and a fit to the  galaxy luminosity function (dotted black) obtained by  \citet{2007ApJ...670..928B}. 
For both QSOs and galaxies observational constraints from a range of studies are shown according to the legend in the plot 
(\citealt{2015AA...578A..83G}, \citealt{2009AJ....138..305J}, \citealt{2015ApJ...810...71F} and \citealt{2015ApJ...803...34B}).
The green cross shows luminosity and space density of the two bright sources in our model 
of the \lya opacity PDF at $z=5.8$ in paper I.}
    \label{lF_set}
  \end{center}
 \end{figure}

We expand here on this by populating DM haloes in the Millennium simulation 
with bright sources (AGN) drawn from a model luminosity function. The latter
is used to describe the UV luminosity of QSOs and their contribution to the ionizing emissivity. Its choice is guided by the observed 
space density of QSOs at these redshifts (see Fig. \ref{lF_set}). We sort halos by 
mass from the most massive to the least massive object and assign ionizing 
luminosities to them which we draw from the model luminosity function in rank order (the most luminous ionizing source is placed in the most massive halo and so on).   
We then  compute the photoionization rate $\Gamma$ due to these bright sources at 
every position in the $500^3$ $\mathrm{(Mpc/h)^3}$ 
volume with  the simple attenuation model used  by \citet{2015MNRAS.447.3402B} (their galaxy UVB model of section 4.2, 
but see  also \citealt{2011MNRAS.414..241B} and \citealt{2013PhRvD..88d3502V}).

In the calculation, we assume a spectral energy distribution appropriate for AGN for the bright sources of the form (see \citealt{2001AJ....122..549V} and \citealt{2002ApJ...565..773T}), 
\begin{equation}
L_{\nu} \propto \left\{\begin{array}{ccc}
 \nu^{-0.44} & \mbox{ if $\lambda > 1300 \, \mathrm{\AA}$ }\\
 \nu^{-1.57} & \mbox{ if $\lambda < 1300 \, \mathrm{\AA}$}
 \end{array}\right.
\end{equation}
This allows us to convert between the $L_{\nu}(1450)$ luminosities probed by observations and the ionizing luminosities governing the ionization state of the IGM.

At each spatial position, we  compute the specific intensity of the ionizing background between 1 and 4 Ry 
by summing over the contribution from each bright source,
\begin{equation}
J(\mathbf{r},\nu)=\frac{1}{4 \pi}\sum_{i=1}^N \frac{L_i(\mathbf{r}_i,\nu)}{4\pi|\mathbf{r}_i-\mathbf{r}|^2}e^{\frac{-|\mathbf{r}_i-\mathbf{r}|}{ \lambda_{\mathrm{mfp}}^{912}}\left(\frac{\nu}{\nu_{912}}\right)^{-3(\beta-1)}}
\label{back_intensity}
\end{equation}

where, $\nu_{912}$ is the frequency at the HI ionizing edge, and $\beta =1.3$ is
the slope of the HI column density distribution, which gives
the dependence of mean free path on frequency  (see also \citealt{2010ApJ...721.1448S} and \citealt{2013MNRAS.436.1023B}).
The sum in equation \ref{back_intensity} is performed for all sources within the simulation box and their periodically replicated images. 
In the following we consider different constant values of the mean free path $\lambda_{\mathrm{mfp}}^{912}$. Later on we will allow it to spatially vary. 
The HI photoionization rate due to bright sources is computed as

\begin{equation}
 \Gamma(\mathbf{r})=4\pi \int_{\nu_{912}}^{4\nu_{912}}\frac{\mathrm{d}\nu}{h\nu}J(\mathbf{r},\nu)\sigma_{\mathrm{HI}}(\nu),
\label{gammaAGN}
\end{equation}

where $\sigma_{\mathrm{HI}}(\nu)$ is the photoionization cross-section (calculated from the fit of \citealt{1997MNRAS.292...27H}).

As in paper I we then have investigated combining  the photoionization rates  due to bright sources in our model  
with the ionizing UV background  due to the much more numerous  galaxies driving reionization in our ATON simulations.
We then calculated the expected effect of the rare bright sources on the \lya opacity PDF in a model where both galaxies and QSOs 
contribute to the ionizing background. 

We thereby combine the contributions from galaxies and QSOs as follows. 

\begin{itemize}
\item We randomly choose a line-of-sight through  the 500$^3$ $({\rm cMpc/h})^3$ volume (along one of the principal axis)
         for which we have modelled  the contribution of bright sources to the photoionization rate $\mathrm{\Gamma_{\rm QSO}^{\rm fiducial}}$.  
\item  We concatenate  three randomly  selected  line-of-sights from the `512-20-good\_l' full radiative  
        transfer simulation and place them along the line-of-sight through the bright source model and call the
        photoionization rates  due to the sources in the radiative transfer simulation  $\mathrm{\Gamma_{\rm gal}^{\rm fiducial}}$ and choose a mean free path.
 \item We calculate the combined photoionization rates  along  the line-of sight, $\Gamma_{\mathrm{gal \, + \, QSO}}=
\Gamma_{\mathrm{gal}}+ \Gamma_{\mathrm{QSO}} = 
\beta_{\rm gal}\Gamma_{\mathrm{gal}}^{\rm fiducial}+ \beta_{\rm QSO}\Gamma_{\mathrm{QSO}}^{\rm fiducial}$ .
 \end{itemize}

As we will see later the contribution from galaxies is thereby poorly constrained  by the  PDF of the \lya opacity as 
long as their is a significant contribution from QSOs. We will thus show results for two models bracketing the contribution 
of galaxies, one where we take into account all luminosities in  our "fiducial"  `512-20-good\_l' model with a factor
 $\beta_{\rm gal} = 1.0$. We also consider models with a slightly larger contribution from galaxies,  $\beta_{\rm gal} = 1.25$.
We then determine the value of $\beta_{\rm QSO}$ by which we need  to rescale  the luminosities  of the bright sources
(QSOs) to match the PDF of the \lya effective optical depth. 
For  each  model for the spatial fluctuations of the photoionization rate we  calculate 5000 mock \lya spectra 
for 50 cMpc/h chunks, i.e. using the same chunk size as in the observed sample of \citet{2015MNRAS.447.3402B}, based on the density, temperature, 
and peculiar velocity fields from the `512-20-good\_l' model (as e.g. described in \citealt{1998MNRAS.301..478T}).

\begin{figure}
  \begin{center}
     \includegraphics[width=9cm,height=7cm]{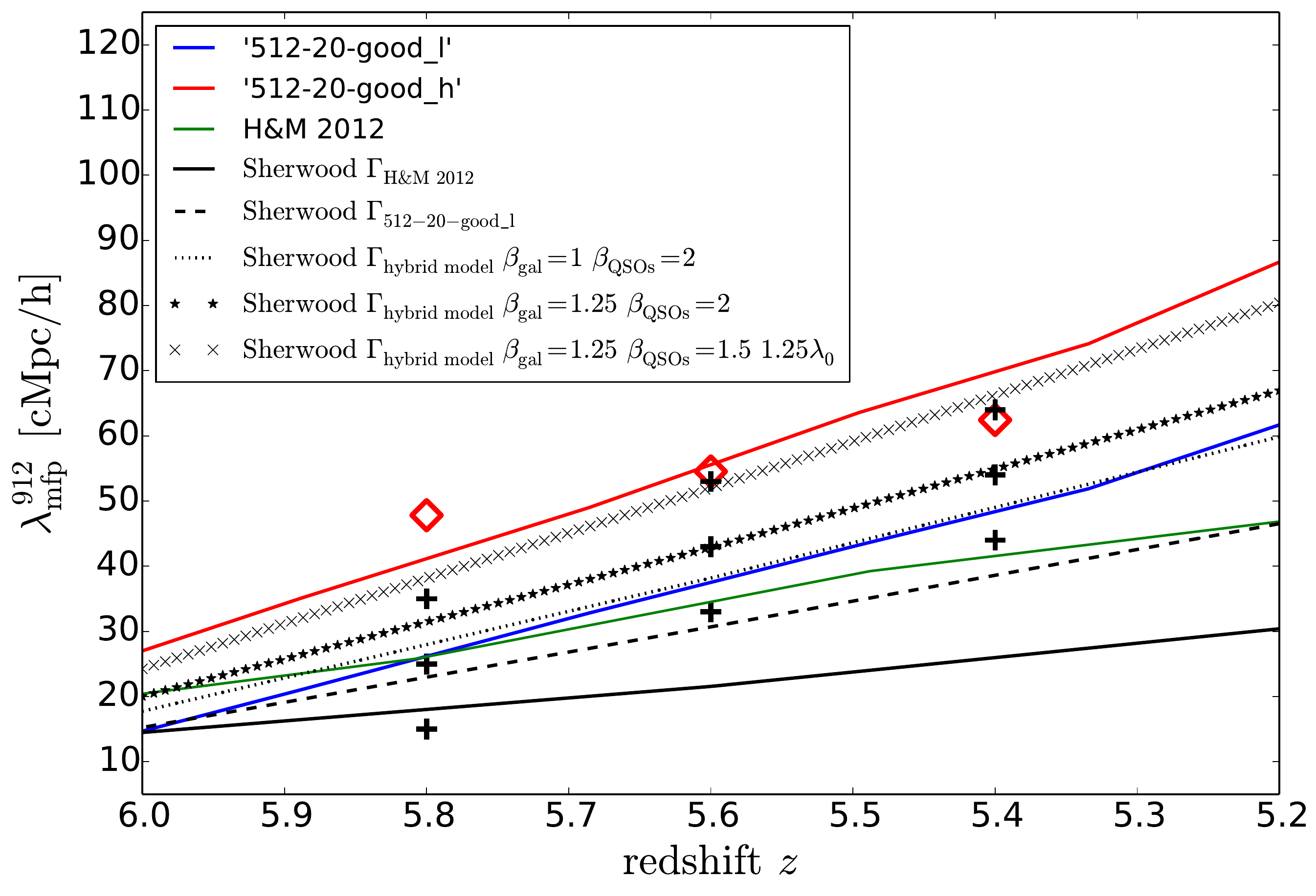} 
  \caption{The range of mean free path values assumed in our models. 
The solid blue and red lines show respectively the evolution of the mean fee path $\lambda_{\mathrm{mfp}}^{912}$ 
in the reference models ``512-20-good\_l'' and ``512-20-good\_h'' as a function of redshift.
The green line shows the evolution of $\lambda_{\mathrm{mfp}}^{912}$ in the \citet{2012ApJ...746..125H} model.
The different black crosses show the values assumed in our different models with fixed mean free path.
The open red diamonds show the extrapolation of the mean free path measured by \citet{2014MNRAS.445.1745W} to higher redshift.
The different black curves shows the evolution of the mean free path in the Sherwood simulation for different 
values of the mean photoionization rate in the three redshift bins : 
assuming the \citet{2012ApJ...746..125H} evolution (solid), assuming the mean $\Gamma$ in our fiducial radiative transfer model ``512-20-good\_l'' (dashed) 
and using the converged value of $\left<\Gamma\right>$ in our hybrid model (dotted).}
    \label{mfp_set}
  \end{center}
 \end{figure}

\begin{figure*}
   \begin{center}
      \includegraphics[width=\textwidth,height=\textheight,keepaspectratio]{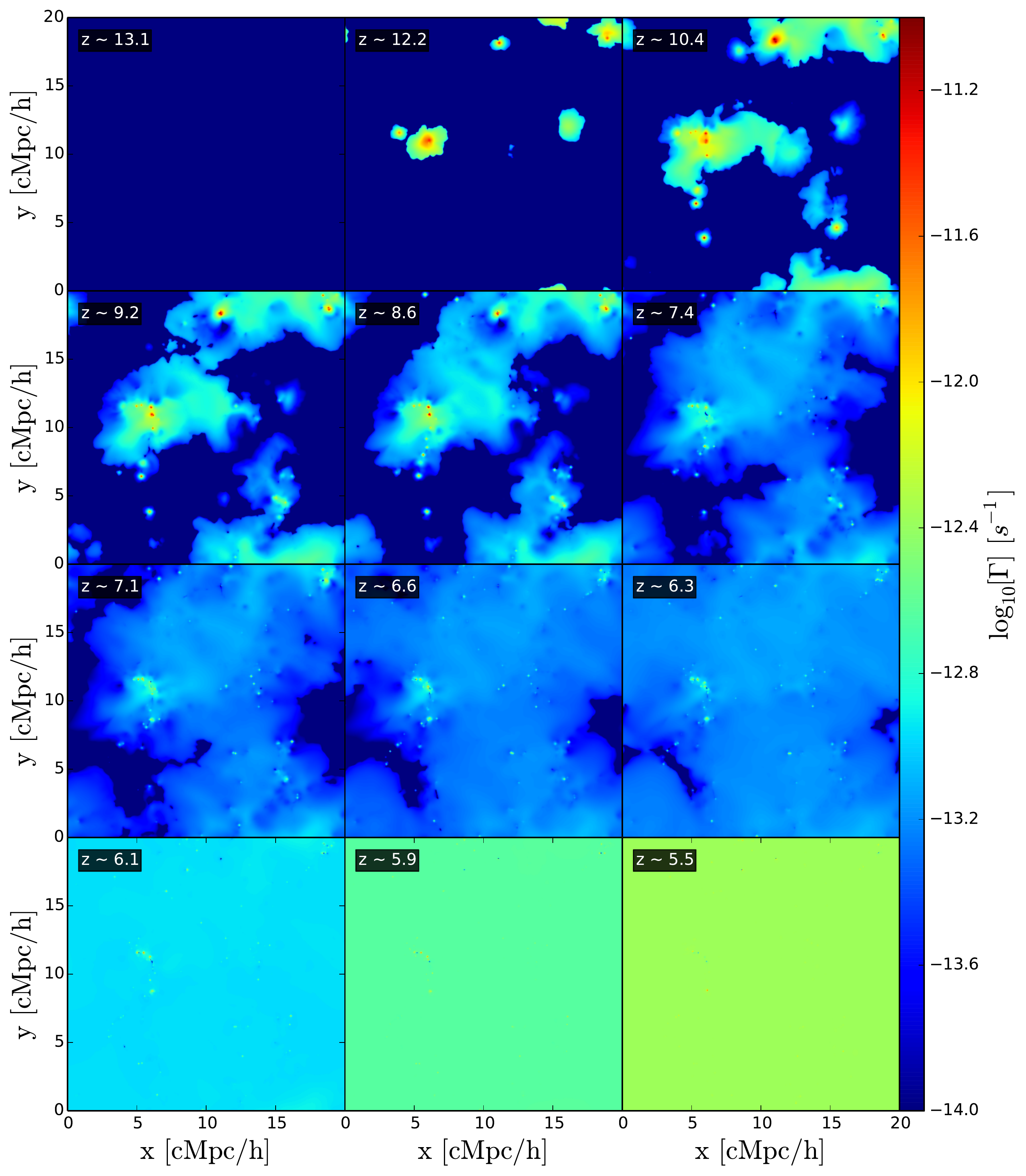}    
  \caption{Redshift evolution of the spatial distribution of $\Gamma$ in our fiducial `512-20-good\_l' full radiative transfer 
  simulation  in a slice  of thickness  39.0625 comoving kpc/h.
}
    \label{mapgamma10redshift}
  \end{center}
 \end{figure*}

\subsection{Modelling UVB fluctuations due to bright sources (QSO)}
\label{parameter_space}

\subsubsection{The luminosity function of the bright sources (QSO)}
\label{LFunctionsubsection}

We want to investigate here in particular whether  QSOs drawn from  a luminosity function
at $z>5$ consistent with the (still rather sparse) data at this redshift can produce sufficiently large spatial 
UVB fluctuations on large scales to reproduce the B2015 measurements. The recent observational 
constraints on the observed   luminosity function of \citet{2015AA...578A..83G}
can be reasonably fit with  a two-power law  (see figure \ref{lF_set}),   

\begin{equation}
 \phi=\frac{\phi^*}{10^{0.4(M_{*}-M)(\beta-1)}+10^{0.4(M_{*}-M)(\gamma-1)}},
\label{power_law_fit}
\end{equation}

with parameters $\mathrm{M_{*}}$, $\beta$ and $\gamma$ as given 
in  table \ref{tab1}. 
We sample the luminosity function from the brightest  luminosity present 
in our $500^3$ $\mathrm{(cMpc/h)^3}$ volume which corresponds to $\mathrm{M_{AB}}\sim -27$ 
and we adopt a lower limit of $\mathrm{M_{AB}} = -22$ for the faintest QSOs and assume that the 
other QSOs with a fainter luminosity are part of the galaxy population. This results in 1296 QSOs 
in the  $500^3$ $\mathrm{(cMpc/h)^3}$ volume.

\begin{figure*}
   \begin{center}
     \includegraphics[width=\textwidth,height=\textheight,keepaspectratio]{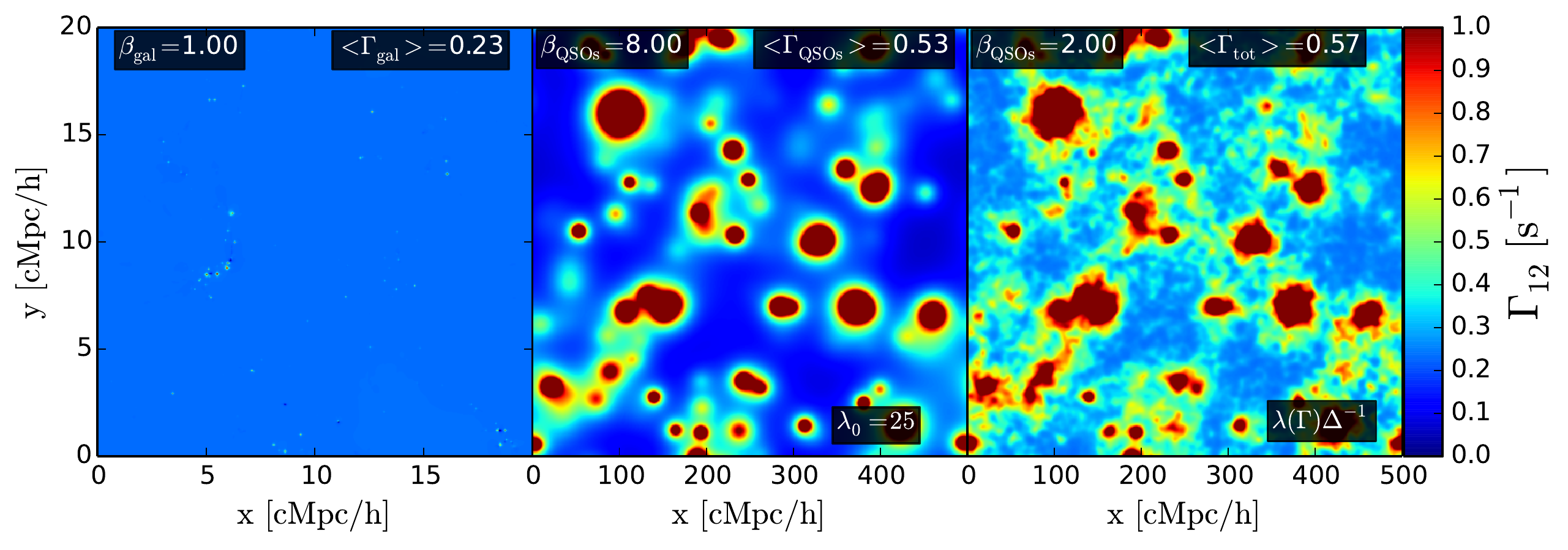} 
  \caption{
Left : Spacial distribution of the photoionization rate $\Gamma_{12}$ in a slice of 39.0625 comoving kpc/h thickness through our fiducial 20 Mpc/h 
radiative transfer simulation  (`512-20-good\_l') at redshift $z\sim 5.8 $ (galaxies only). 
Middle: Spatial distribution of the photoionization rate ($\Gamma_{12}$) due to QSOs  in a slice of of 976.5625 comoving kpc/h through 
the Millennium simulation  at redshift $z\sim 5.8 $ for a constant mean free path of $\mathrm{\lambda_0=25}$ cMpc/h.
Right : Spatial distribution of the  photoionization rate due to QSOs plus galaxies  ($\Gamma_{12}$) for a $\mathrm{\Gamma}$ dependant 
parametrization of the mean free path $\mathrm{\lambda(\Gamma)}$ according to equation \ref{furlanet_gamma} 
with $\mathrm{\lambda_{mfp} \propto \Delta^{-1}}$.
The mean photoionization rate $\mathrm{\left<\Gamma \right>}$  reported in the upper right corner of the left panel 
is that due to galaxies only in our fiducial full radiative transfer simulation and that in the middle panel is that due to the additional 
QSOs contribution $\mathrm{\beta_{QSOs}}$  necessary to match the \lya opacity PDF with a fixed mean free path  of path of $\mathrm{\lambda_0=25}$
Mpc/h.}
    \label{map_combined_gamma}
  \end{center}
 \end{figure*}

\subsubsection{The mean free path of ionizing photons}
\label{mfpsubsection}

As we add the contribution of the UV background by QSO in a post-processing step,  
the next important parameter that we have to choose for the modelling of the spatial fluctuations of the ionizing UV background is the 
mean free path of ionizing photons due to QSOs.  The solid blue and red curves in Figure \ref{mfp_set} show the evolution of the mean 
free path in our fiducial `512-20-good\_h' and  `512-20-good\_l'      
full radiative transfer runs  while  the green fitted curve shows that assumed in the HM2012 UV background model (their best guess of 
extrapolating   the mean free path observed at lower redshift to higher redshift).  
In the redshift range we are interested in here, $z=5.4-5.8$, the mean free path of ionizing photons 
is still rather uncertain and we have first explored a range of fixed values  of mean free path 
spanning the range between that assumed by HM2012 and our fiducial 
radiative transfer simulations  as shown by the small crosses in Figure \ref{mfp_set} and summarised 
in Table \ref{tab2}.

The highest values explored in each redshift bin are comparable with the extrapolation of 
measurements of the mean free path by \citet{2014MNRAS.445.1745W} at lower redshift. 
The only exception is z=5.8 where the values we consider are a bit lower.
At this redshift we are approaching the epoch of reionization where the mean free path is highly uncertain and could potentially strongly diverge from the extrapolation. 
The somewhat lower values we use are thus not implausible.
        
Assuming a fixed mean free path  for the ionizing radiation of QSOs is obviously a rather poor approximation as 
the large UV fluctuations due to QSOs will also result in large fluctuations of the mean free path. 
 \citet{2016MNRAS.460.1328D} based on the model of \citet{2000ApJ...530....1M}  and the work of \citet{2011ApJ...743...82M} 
argued that the mean free path in the post-reionization Universe should depend as a simple power law on the 
photoionization rate and density,

\begin{equation}
 \mathrm{\lambda_{mfp}(\Gamma)=\lambda_0(\Gamma/\Gamma_0)^{2/3}\Delta^{-\gamma}}.
\label{furlanet_gamma}
\end{equation}

Here $\mathrm{\Delta}$ is the overdensity in the simulation cells and $\mathrm{\gamma}$ sets the power law dependence on the overdensity for the mean free path.
\citet{2016MNRAS.460.1328D} choose a value of $\mathrm{\gamma=1}$ and we will investigate 
this model  here as well as a model with a somewhat shallower dependence on overdensity (see appendix \ref{lambda_mfp_delta}). 
 As $\Gamma$ itself will depend on the ionizing emissivity in these models as well as the mean 
free path, equation \ref{gammaAGN} has then to be solved iteratively. 

For the normalisation of the mean free path  in  equation \ref{furlanet_gamma} 
we have chosen  $\mathrm{\lambda_0}$ and $\mathrm{\Gamma_0}$  such 
that we reproduce the mean free path in our numerical simulation in the limit of 
a spatially constant UV background. 

To this end we have measured the mean free path in the 40-2048-ps13 run of the Sherwood Simulation Suite \citep{2016arXiv160503462B} 
using the method described in Appendix C of Paper I 
and applying the \citet{2013MNRAS.430.2427R} self-shielding prescription. 
The simulation assumes a \citet{2012ApJ...746..125H} UV background and includes a wide range of galaxy formation physics \citep{2013MNRAS.428.2966P}. 
The mean free path is averaged over many random lines-of-sight. We then adopt the measured mean free path as $\lambda_0$ and the assumed photoionization 
rate from \citet{2012ApJ...746..125H} as $\Gamma_0$.
The corresponding values for $\mathrm{\lambda_0}$ and $\mathrm{\Gamma_0}$ are reported in Table \ref{tab2} for the three redshift bins considered.
We have also tested by rescaling the ionized fractions to a different photoionization rate and by selecting lines of sight with a suitable smoothed overdensity 
at the starting point that Eq. \ref{furlanet_gamma} faithfully describes the scaling of the mean free path with $\Gamma$ and $\Delta$. 
A value of $\gamma \approx 0.4$ is preferred by this comparison, but larger values of $\gamma$ up to unity are obtained when measuring 
the mean absorption over short path lengths.

Note that in our $\Gamma$ dependent model for the mean free path we have also iteratively accounted for the expected dependence 
of the photoionization rate due to galaxies on the mean free path by modulating the photoionization rate predicted by our 
full radiative transfer simulation accordingly,
\begin{equation}
 \mathrm{\Gamma_{\rm gal}^{\rm fiducial}(\Gamma_{\rm gal+QSO})=\Gamma_{gal\,RT}\frac{\lambda_{mfp}(\Gamma_{QSO+gal})}{\lambda_{mfp}(\Gamma_{gal\, RT})}},
\end{equation}

where  $\mathrm{\Gamma_{gal\,RT}}$ is the  photoionization rate due to galaxies in the radiative transfer simulation at a given location 
and we have assumed the galaxy mean free path to scale with photoionization rate and over-density 
as in equation \ref{furlanet_gamma}.


\begin{figure*}
   \begin{center}
      \includegraphics[width=\textwidth,height=\textheight,keepaspectratio]{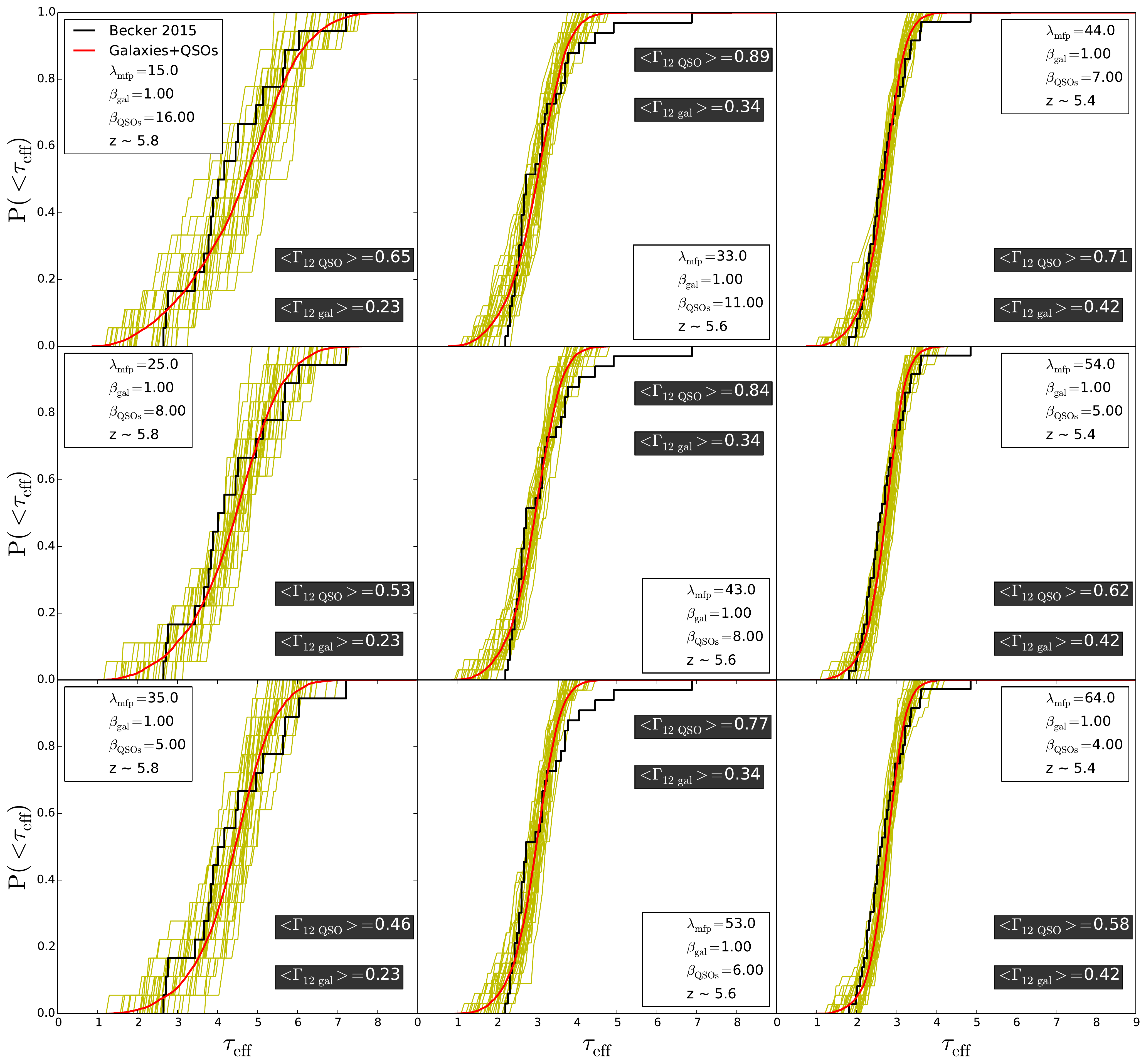}   
  \caption{PDF of $\mathrm{\tau_{eff}}$ for the rescaled luminosity functions with the  same shape as the fit in  \citealt{2015AA...578A..83G})  
and different fixed mean free paths 
at the three different redshifts (first column: z $\sim $ 5.8, second column: z $\sim $ 5.6, third column: z $\sim $ 5.4).
The  different constant mean free paths considered at the three redshifts are (z $\sim $ 5.8: 15,25,35 cMpc/h, z $\sim $ 5.6: 33,43,53 cMpc/h,
 z $\sim $ 5.4: 44,54,64 cMpc/h) 
as discussed in Section \ref{mfpsubsection}.}
    \label{PDF_comparison1}
  \end{center}
 \end{figure*}

\section{Results}
\label{results}

\subsection{Spatial fluctuations of  the photoionization rate distribution in the full radiative transfer simulations}
\label{gammaRTdistrib}

First, we look at the spatial distribution of the photoionization rate $\Gamma$ in our full radiative transfer simulations at different redshifts.
Figure \ref{mapgamma10redshift} shows  the redshift evolution for our fiducial `512-20-good\_l' simulation. As expected before overlap of the 
HII regions there are large spatial variations but as discussed in paper I these damp out very quickly after overlap of the HII regions which occurs 
at $z=6.5-6$ (see Fig. 11 of paper I for the redshift evolution of the PDF of $\mathrm{\Gamma_{12}}$). At z=5.9 the photoionization rate in our fiducial simulation is already  very homogeneous, 
but see \citealt{2016MNRAS.460.1328D} for a discussion how the rather small volume covered by our simulations may affect 
the amplitude and scale of spatial fluctuations of the photoionization rate. Note, however, that we did not see any large effect 
of the box-size of our simulations in paper I.

\subsection{Spatial  fluctuations of the photoionization rate  in our combined UVB model of (faint) galaxies and  QSOs}
\label{gammatoydistrib}

\begin{figure*}
   \begin{center}
      \includegraphics[width=\textwidth,height=\textheight,keepaspectratio]{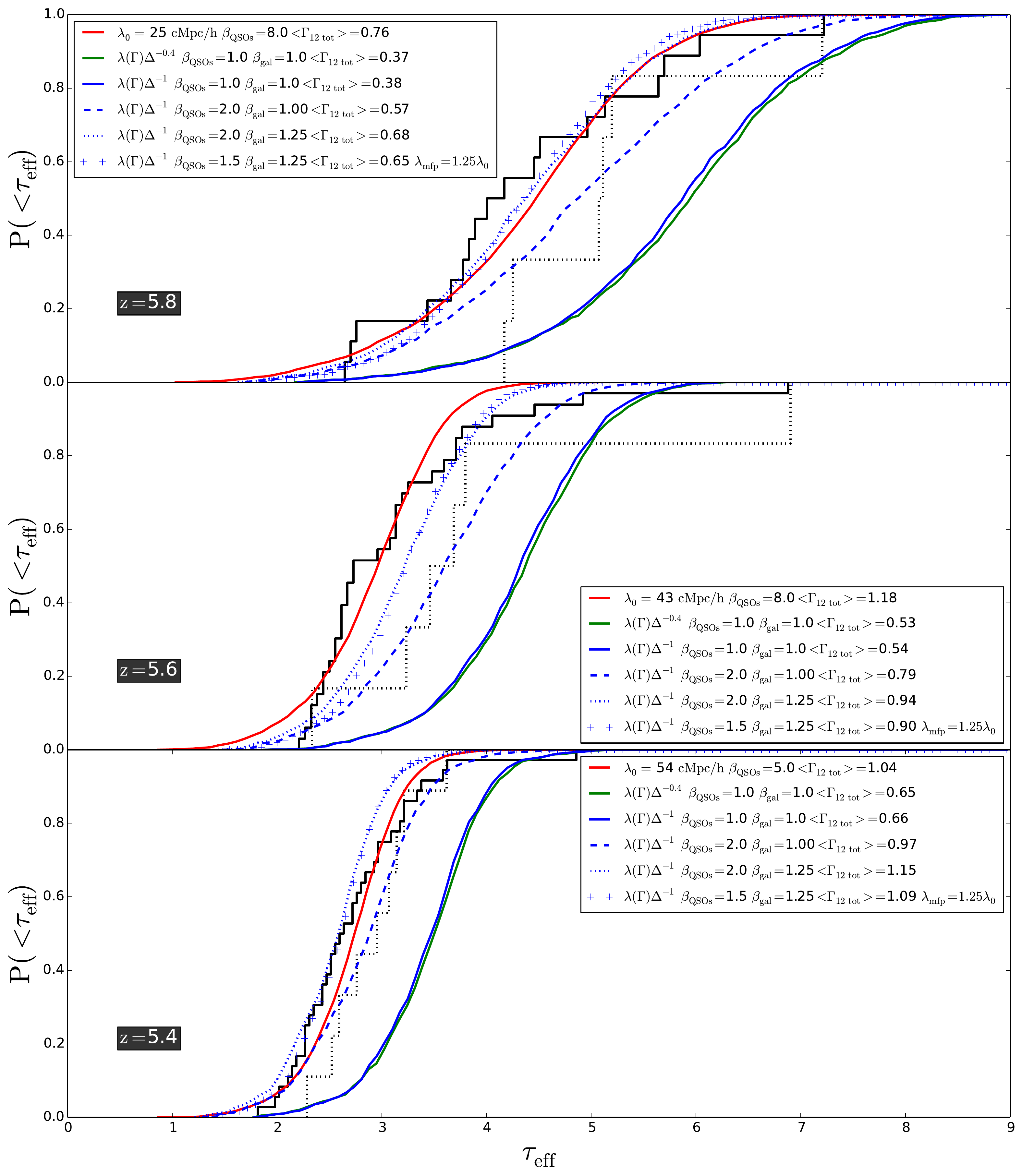}   
  \caption{PDF of $\mathrm{\tau_{eff}}$ for models with  the (rescaled) luminosity function of  \citealt{2015AA...578A..83G}  at the three redshifts  with values $\mathrm{\beta_{gal}}$ and $\mathrm{\beta_{QSOs}}$ as indicated in the plots.
The red curves  shows the case with a constant mean free path $\mathrm{\lambda_{mfp} = \lambda_0}$ while the other solid colored curves show  
cases with a varying mean free path adopting the different parametrization chosen in equation \ref{furlanet_gamma}.
The black solid step function shows the data from \citet{2015MNRAS.447.3402B} based on their own sample of QSO spectra 
combined with the sample of \citet{2006AJ....132..117F}   while the dotted  black  step function 
is for  the \citet{2015MNRAS.447.3402B} sample only.}
    \label{PDF_comparison2}
  \end{center}
 \end{figure*}

Figure \ref{map_combined_gamma} shows maps of the photoionization rate $\mathrm{\Gamma_{gal}}$ and $\mathrm{\Gamma_{QSO}}$ at $z\sim5.8$ assuming a fixed mean free path.  
The maps shown are the ones that best match the observed cumulative \lya effective optical depth PDF as explained in  section \ref{PDFteff}.

For the fit to the QSO     luminosity function as in  \citet{2015AA...578A..83G} the luminosities need to be rescaled by a factor 
 $\mathrm{\beta_{QSO} = (16,8,5)}$
in order to match the observed \lya effective optical depth PDF at redshift z=5.8 for our models with  mean free path of 
(15,25,35) cMpc/h.
These rather high values of $\mathrm{\beta_{QSO}}$ are due to the fact that our assumed mean free
path is lower than the (mean) distance between the  QSOs in our model.  As already
discussed neglecting the effect of the QSOs on the mean free path is a bad
approximation.
As we will see the required luminosities to match the PDF are significantly lower  for our more realistic  models with a $\Gamma$ dependent mean free path.   
The upper right panel shows the effect of introducing  a $\Gamma$ dependent mean free path as described in section \ref{mfpsubsection}.
In this case the value of  $\mathrm{\beta_{QSOs}}$  decreases to $\sim 1.5-2$  and the luminosity function required to 
reproduce the observed opacity fluctuations gets close to that observed by  \citet{2015AA...578A..83G}.  As we will discuss 
later this agreement should improve further if the expected increase of the temperature in the proximity of the QSO
due to ionization of HeII would be taken into account. This is currently neglected in our simulations.

\subsection{Opacity fluctuations in the \lya forest  in the combined UVB model of (faint) galaxies and QSOs}
\label{PDFteff}

Here, we present in Fig. \ref{PDF_comparison1} the  cumulative \lya optical depth PDFs of our mock absorption spectra for the  different models.
We choose to show the cumulative rather than the differential PDFs as the former were discussed in the original observational paper of \citet{2015MNRAS.447.3402B}.
We have shown in paper I that for the case of a galaxy only model (i.e. $\mathrm{\beta_{QSOs}}=0$) the \lya optical depth PDFs are too steep and cannot explain the
observed scatter.
As already mentioned in the last section we are able to match the $\mathrm{\tau_{eff}}$ PDF in all redshift bins with our combined model of galaxies and QSOs.
In each panel  the yellow curves show 30 different realizations of the PDF computed with the same path length as 
in the \citet{2015MNRAS.447.3402B} sample in the corresponding redshift bins.
In every model, the scatter of the PDF brackets the observations reasonably except perhaps the high opacity point in 
the redshift bin z=5.6.  We should note here that these large optical depths can all be traced to the very long 
($110  h^{-1}$ comoving Mpc)  trough in ULAS J0148+0600.  If we allow for a rescaling of the luminosities, matching 
the PDF is not very difficult, but  the amount by which we have to rescale the photoionization rate $\Gamma$ of the QSO 
counterpart in order to get agreement  with the observations is rather  different from one model to another. In each panel 
we show  the $\mathrm{\beta_{QSOs}}$ values we need to match the PDF for a fixed $\mathrm{\beta_{gal}} =1$. 
The $\mathrm{\beta_{QSOs}}$ values were found by testing qualitatively how well different values reproduce the observed opacity PDFs.

In Fig. \ref{PDF_comparison1}, we show the effect of varying the assumed (spatially constant) mean free path. 
We can see that changing the mean free path moderately changes the shape of the PDF in all the redshift bins, 
especially for the two lower redshifts z=5.6 and z=5.4. With increasing mean free path the opacity PDF becomes narrower, 
while  the ionizing emissivity  and to a smaller extent the contribution to the photoionization rate from QSOs required 
to match the observed PDF decreases. Similar  conclusions have been reported by \citet{2015MNRAS.447.3402B}.
Lowering the mean free path leads to larger photoionization rate fluctuations as the low density IGM would 
see less and less ionizing photons.

Let us now  have a look at the effect of introducing a  spatially varying mean free path. 
In figure \ref{PDF_comparison2}, we compare the \lya effective optical depth PDF for the case of a constant mean free path  
and with a set of $\Gamma$ dependant mean free path parametrization as described in section \ref{mfpsubsection}. 
As already mentioned, the value of $\mathrm{\beta_{QSOs}}$ decreases to $1.5-2$
with the $\Gamma$ dependant mean free path parametrization at z=5.8 
if we assume the mean free path to depend on overdensity as  $\mathrm{\Delta^{-1}}$.
Large values of the mean free path close to ionized regions have the effect of increasing the  $\mathrm{\left<\Gamma_{QSOs}\right>}$ values.
Therefore, for a given luminosity function, adopting a photoionization dependant mean free path parametrization in the combined UVB model leads to a lower value of $\mathrm{\beta_{QSOs}}$
required to generate photoionization rate fluctuations that match the PDF of $\mathrm{\tau_{eff}}$ .

A weaker dependence of the mean free path on overdensity as  $\mathrm{\Delta^{-0.4}}$ as suggested by 
our measurements of the mean free path in the Sherwood simulation 
leads to similar values of $\mathrm{\beta_{QSOs}}$ needed to match the $\mathrm{\tau_{eff}}$ PDF.
The overdensity dependence of the mean free path seems therefore to play only a minor role for the resulting UV background and the related $\mathrm{\tau_{eff}}$ PDF. We have also tested  different smoothing scales for the overdensity 
field $\mathrm{\Delta}$ measured from the  Millennium simulation. This also changes our results  very little (see Appendix \ref{lambda_mfp_delta}).

We should, however, note that for models with  the original luminosity function of \citet{2015AA...578A..83G} ($\mathrm{\beta_{QSOs}}=1$), 
the $\mathrm{\tau_{eff}}$ values are somewhat high compared to the data of \citet{2015MNRAS.447.3402B}. We therefore had a closer look 
at the data of \citet{2015MNRAS.447.3402B} and noted that there is a noticeable  offset between the PDF reported 
by  \citet{2015MNRAS.447.3402B} based on their own sample of QSO spectra  combined with the sample of \citet{2006AJ....132..117F}   
compared to the PDF  for  the \citet{2015MNRAS.447.3402B} 
sample only (shown as the dotted step function in Fig.  \ref{PDF_comparison2}).  
The latter is indeed preferring models with a somewhat lower photoionization rate. This offset is somewhat larger 
than could be due to statistical fluctuations for samples of this size. We convinced ourselves that at least part of the difference
could be due to a more conservative cleaning of \citet{2015MNRAS.447.3402B}  of their own sample for regions where the observed 
flux is affected by proximity effects of the observed QSOs (caused by their ionizing flux and to a lesser extend the matter clustering around them). The size of the proximity regions is strongly dependent on the 
assumed photoionization rate due to surrounding galaxies and therefore not straightforward at these redshifts.

\subsection{The lengths of Gunn-Peterson troughs}
\label{GPtrough}

\begin{figure*}
   \begin{center}
      \includegraphics[width=\textwidth,height=\textheight,keepaspectratio]{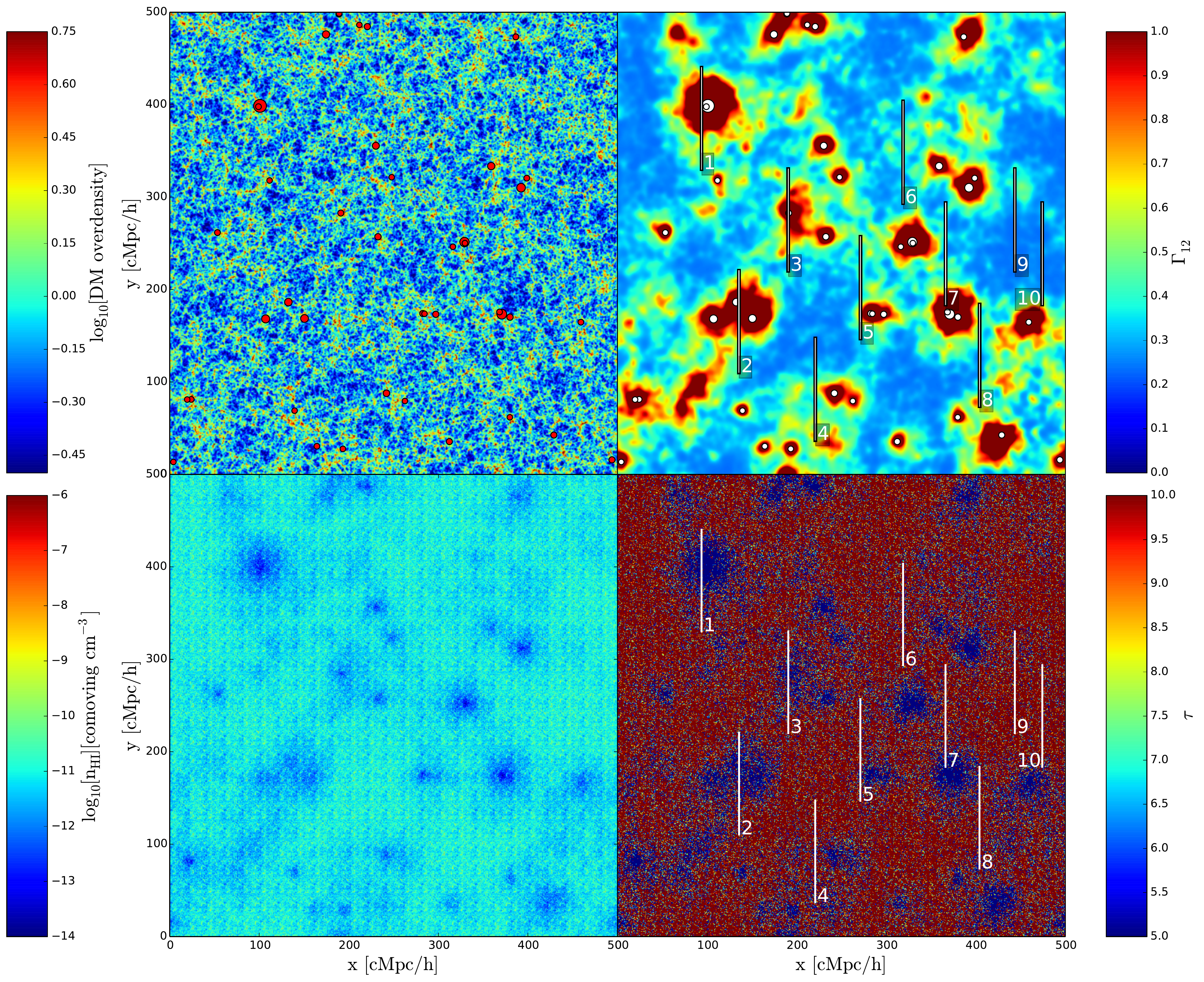}    
  \caption{Top left: Spatial distribution of the dark matter overdensity in the Millennium volume at $z\sim5.8$ in a slice of thickness  976.6 comoving kpc/h. The red circles show the location of the DM haloes hosting the QSOs. 
Top right:  The spatial distribution of the photoionization rate $\Gamma_{12}$ in the volume of the Millennium volume simulation 
for the fit  of the QSO  luminosity function of \citealt{2015AA...578A..83G} with 
the $\Gamma$ dependant mean free path case, $\mathrm{\lambda_{mfp}=\lambda_0(\Gamma/\Gamma_0)^{2/3}\Delta^{-1}}$.
The map has been calculated for  a 512 $\times$ 512 grid and then interpolated on a 4096 $\times$ 4096 grid.
The overdensity field $\mathrm{\Delta}$ from the Millennium simulation has been smoothed  on a 20 comoving Mpc scale with a top-hat filter
to calculate the $\Gamma_{12}$ field in the slice.
Bottom left: The spatial distribution of the neutral hydrogen number density inside the Millennium volume : The density field in the mid-plane slice of the `512-20' RAMSES simulation 
is replicated 25$\times$25 to cover the full size of the Millennium simulation  and has been interpolated on  a 4096 $\times$ 4096 grid.
Bottom right: the spatial distribution of the opacity $\tau$   in the same slice calculated using the interpolated 
4096 $\times$ 4096 grids of the hydrogen number density (bottom left panel), velocity (not shown in the plot), temperature (not shown in the plot) and photoionization rate (top right panel). 
The red and white points respectively in the upper and bottom left panels show the
position of the dark matter haloes assumed as ionizing sources in our model (taken in a slice of $\mathrm{\sim 15 \, cMpc/h}$ thickness around the slice shown). 
The black/white thick lines in the bottom left/right panel show 10 lines-of-sights of 110 Mpc/h 
length along which we compute and show the corresponding spectra in figure \ref{example_spectra}.
The maps shown are calculated with $\beta_{\rm QSOs}=2$ and  $\beta_{\rm gal}=1$.}  
    \label{various_millenium_map}
  \end{center}
 \end{figure*}

\begin{figure*}
   \begin{center}
      \includegraphics[width=\textwidth,height=\textheight,keepaspectratio]{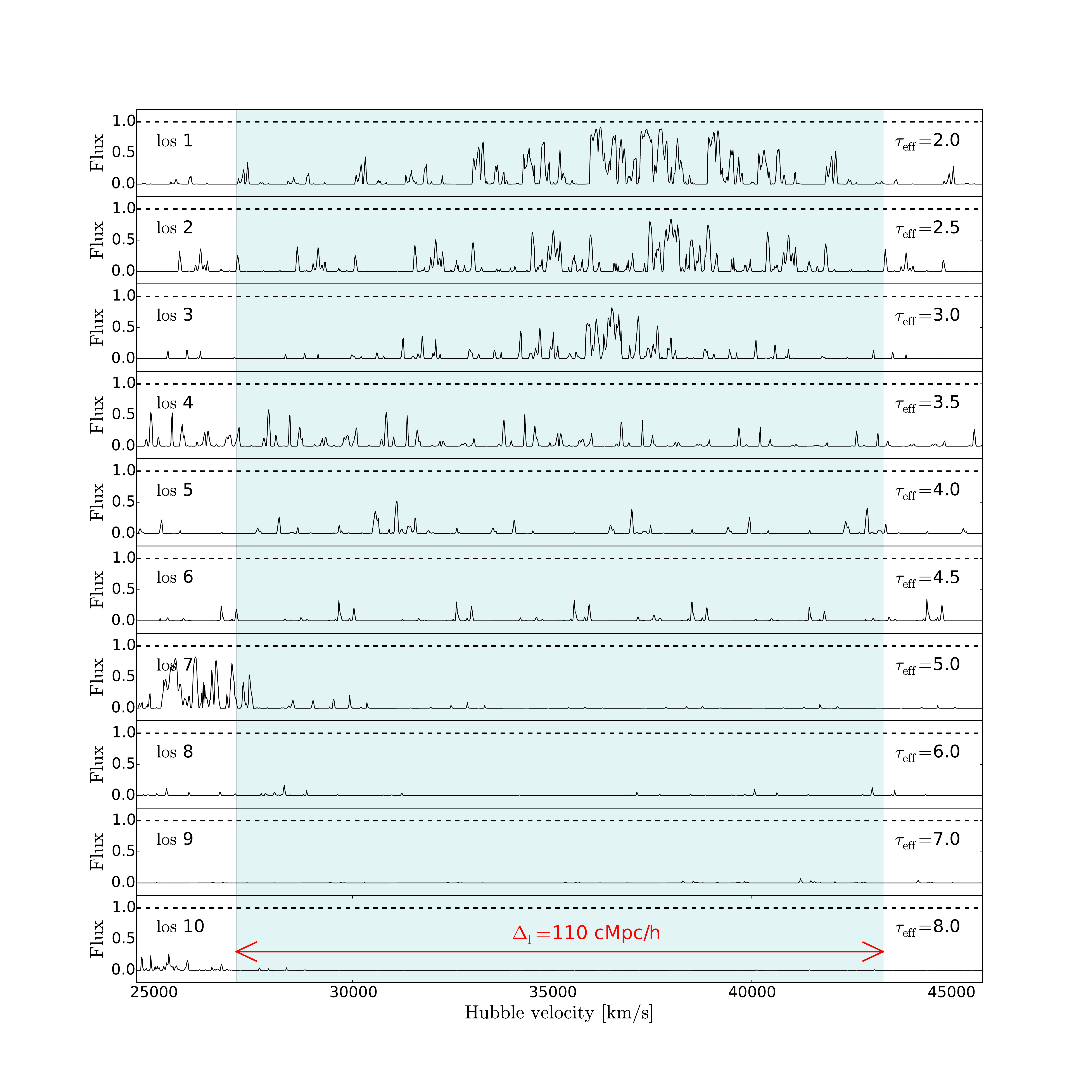}    
  \caption{Example of 10 spectra obtained along one of the principal axis at z=5.8 for our model with the  luminosity function of \citealt{2015AA...578A..83G}) 
and with a varying mean free path : $\mathrm{\lambda_{mfp}=\lambda_0(\Gamma/\Gamma_0)^{2/3}\Delta^{-1}}$.
The overdensity field $\mathrm{\Delta}$ from the Millennium simulation has been smoothed on a scale of 20 comoving Mpc to  calculate the spectra. 
The light blue shaded area represents a comoving size of 110 Mpc/h and the corresponding effective optical depth in that chunk is shown in the lower right of each panel. 
The spectra are ordered from the lowest $\mathrm{\tau_{eff}}$ for these  chunks of 110 Mpc/h to the highest from top to bottom.
The  lines-of-sight corresponds to those shown in figure \ref{various_millenium_map} as labelled  there and in the  upper left corner of each panel here.
The spectra are calculated with $\beta_{\rm QSOs}=2$ and  $\beta_{\rm gal}=1$ as in Figure. \ref{various_millenium_map}.}
    \label{example_spectra}
  \end{center}
 \end{figure*}

\begin{figure*}
  \begin{center}
     \includegraphics[width=\textwidth,height=\textheight,keepaspectratio]{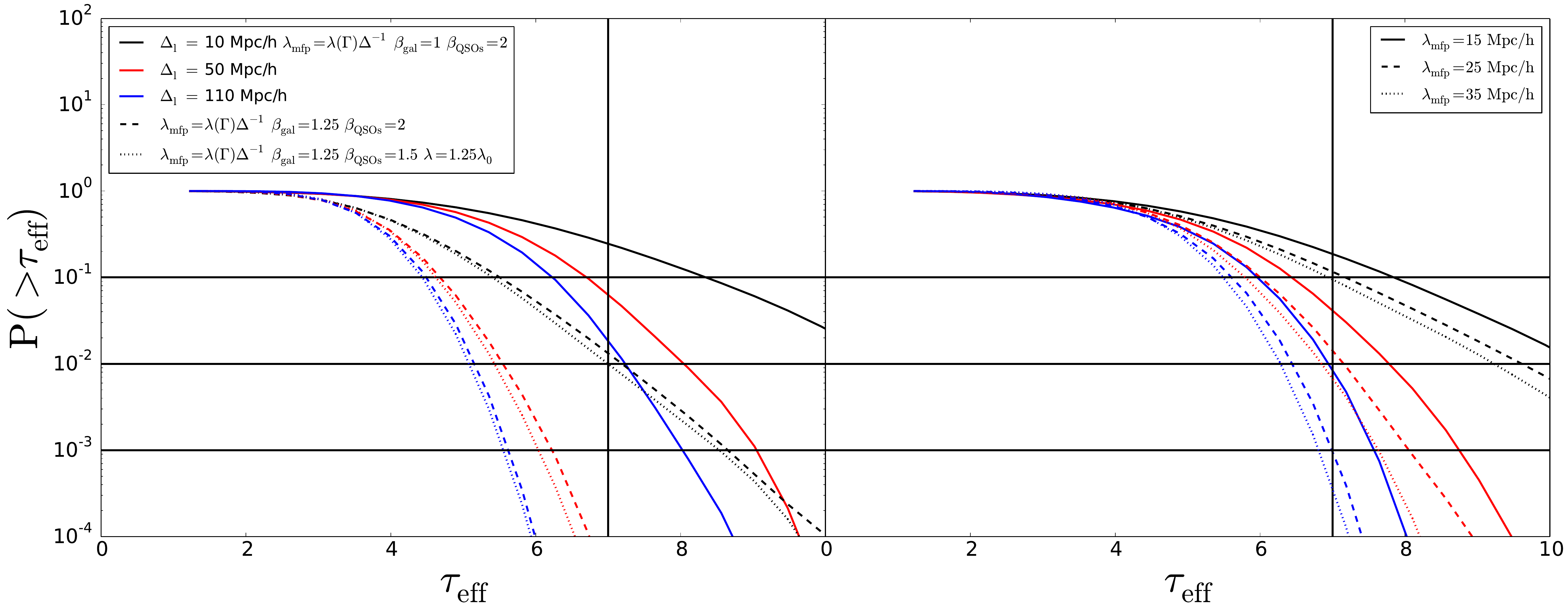} 
  \caption{The cumulative effective optical depth PDF $\mathrm{P(>\tau_{eff})}$ 
for chunks of spectra of different length  ($\mathrm{\Delta_l=10,50,110}$ Mpc/h in black, red and blue respectively) at $z = 5.8$ for different assumptions of the 
mean free path. 
The solid lines in the left panel stand for the model with $\beta_{\rm QSOs}=2$ and  $\beta_{\rm gal}=1$, the dashed lines for $\beta_{\rm QSOs}=2$ and  $\beta_{\rm gal}=1.25$ 
and the dotted lines for $\beta_{\rm QSOs}=1.5$ and  $\beta_{\rm gal}=1.25$ and 1.25 times the original value of the mean free path.
The vertical black line corresponds to a value of $\mathrm{\tau_{eff}=7}$  
as measured for  the large trough of 110 Mpc/h in the spectrum of ULAS J0148+0600  by  \citet{2015MNRAS.447.3402B}.
 The  black horizontal lines show 10\%, 1\% and 0.1\%.}.
   \label{PDF_teff_troughs}
  \end{center}
 \end{figure*}

As we have seen reproducing the large opacity fluctuations reported by B2015 on large scales  ($\ga 50  h^{-1}$ comoving Mpc)
is difficult and the very long  $(110  h^{-1}$ comoving Mpc) high opacity trough in ULAS J0148+0600 is particularly challenging.
The box sizes of our high-resolution hydro-simulation are significantly smaller than this, but  the scale of the opacity fluctuation
in our model is set by the UV fluctuations due to the proximity effect which we model as before with the 
$500^3$ $\mathrm{(Mpc/h)^3}$ Millennium simulation.

In order to see if we are able to reproduce such a long high-opacity trough in our hybrid 
simulations we proceeded as follows.

\begin{itemize}
 \item{We choose a random line of sight through the $(500 \, \mathrm{Mpc/h})^3$ Millennium volume along one of the principal axis.}
 \item{We calculate the $\Gamma_{\rm QSO}^{\rm fiducial}$ values due to QSOs along that skewer discretized in $500/20\times512$ cells with equation \ref{gammaAGN}.}
 \item{We concatenate 500/20=25 randomly selected line-of-sights from the `512-20-good\_l' model and place them along  the 
 500 Mpc/h skewer through the   Millennium simulation.}
 \item{We calculate the combined photoionization rates  along  the line-of sight, $\Gamma_{\mathrm{gal \, + \, QSO}}=
\Gamma_{\mathrm{gal}}+ \Gamma_{\mathrm{QSO}} = 
\beta_{\rm gal}\Gamma_{\mathrm{gal}}^{\rm fiducial}+ \beta_{\rm QSO}\Gamma_{\mathrm{QSO}}^{\rm fiducial}$.}
 \item{We repeat this  procedure for 500 random skewers through the  500 Mpc/h  Millennium simulation. 
         From  each of the 500 skewers  we pick randomly five 110 Mpc/h chunks and calculate the effective optical depth 
         from mock absorption spectra, resulting in 2500 estimates of  $\mathrm{\tau_{eff}}$.}
\end{itemize}

We use here again our simulation with the QSO luminosity function of \citet{2015AA...578A..83G} 
at  $z=5.8$ with  $\beta_{\rm gal}=1$ and $\beta_{\rm QSOs}=2$ for the $\Gamma$  dependant mean free path case 
with an  overdensity dependence as $\mathrm{\Delta^{-1}}$.
We note that the overdensity field $\mathrm{\Delta}$ of the Millennium simulation has been smoothed here on 20 comoving Mpc scale 
to calculate the photoionization rate values.

In figure \ref{example_spectra}, 
we show ten examples of our mock spectra for a range of effective optical depth values and in the top and bottom right panels of Fig. \ref{various_millenium_map}
the corresponding line-of-sights are overlaid on maps of the photoionization rate $\mathrm{\Gamma}$ and the optical depth $\mathrm{\tau}$ of our hybrid simulations. 
The top and bottom left panels of Fig. \ref{various_millenium_map} show the DM overdensity of the corresponding slice of the Millennium simulation and the hydrogen density 
in the concatenated high-resolution hydro-simulations, respectively. 
For each mock absorption spectrum in Fig. \ref{example_spectra}, the  $\mathrm{\tau_{eff}}$ computed in a 110 Mpc/h portion of the skewer marked in blue is given. 
We indeed find a few spectra with   $\mathrm{\tau_{eff}}>7$ as measured for the long trough in 
ULAS J0148+0600. 

Fig. \ref{PDF_teff_troughs} shows the cumulative PDF of $\mathrm{\tau_{eff}}$ inside skewers 
computed from the 500 mock \lya spectra. We show the PDF for three different lengths for  which  $\mathrm{\tau_{eff}}$ values are calculated.  
The vertical line in the figure shows $\mathrm{\tau_{eff}}=7$, the value reported  by \citet{2015MNRAS.447.3402B}  for ULAS J0148+0600,  while   the  horizontal lines show fractions of 10, 1 and 0.1\%.

There is a non negligible tail of the PDF at $\mathrm{\tau_{eff}}>7$ for $\Delta_l$=10 and 50 Mpc/h with respectively about 
30\% and 7\% of the line-⁠of-⁠sight chunks that lie above this value for our model with $\beta_{\rm QSOs}=2$ and $\beta_{\rm gal}=1$ .  
Somewhat surprising this is in reasonable agreement with the recent finding of \citet{2014ApJ...793...30G} who found a value of about 3\% of their 
line-of-sights (computed in 40 Mpc/h chunks) 
that have $\mathrm{\tau_{eff}>7}$ at $z\sim5.8$ in their full hydro-RT simulation that do, however, not contain UV fluctuations due to QSOs.  
There is also a small number  of  line-of-sights that have $\mathrm{\tau_{eff}}>7$. For $\mathrm{\Delta_l}$=110 cMpc/h 10\% of chunks of our mock spectra 
have $\mathrm{\tau_{eff} > 6}$ and about 2\% have $\mathrm{\tau_{eff} > 7}$ as in the observed very long trough in ULAS J0148+0600. 
Since the Ly$\alpha$ forest part of the spectrum of a $z\sim 6$ QSO extends about 370 cMpc/h, we predict a $\sim$ 7\% chance of 
it containing such a long absorption trough. The fraction of such spectra drops, however rapidly if we increase $\beta_{\rm gal}$ 
suggesting that reionization by galaxies occurs as late or later as in our fiducial 512-20-good\_l full radiative transfer simulation.

Comparing the individual spectra with the lower panels of Fig. \ref{various_millenium_map} is rather illustrative. 
Individual regions of high transmitted flux can be clearly traced back to the transverse proximity effect due to the 
corresponding line-of-sight passing rather close to a QSO. This is a clear observational prediction of the model but relies 
crucially on the assumed long-life times of the QSO  and  isotropic nature of their emission. 
Unfortunately, failure to find easily the signature of a transverse proximity effect will therefore not necessarily rule out the 
QSO proximity effect as a source of the opacity fluctuations. Instead it will put limits on the duty cycle and angular distribution of the emission of QSOs. 
Interestingly, \citet{2008MNRAS.386..359G} et al have already reported the detection of a transverse proximity effect at $z=5.70$ in the \lya forest part of the spectrum of
SDSSJJ1148+5215.

\section{Discussion}
\label{discussion}

\subsection{Are there a sufficient number of  QSOs ?}

As we had seen the required contribution by QSOs to the ionizing emissivity is model dependent
and depends in particular on the modelling of the mean free path for ionizing photons. This is still
very difficult to do self-consistently as full radiative transfer simulations including QSO with sufficient 
dynamic  range to properly resolve the sinks of reionization/Lyman-Limit-Systems are not yet possible. For our most 
realistic model of a $\Gamma$ dependent mean free path calibrated with a high-dynamic range 
hydro-dynamical simulation, QSOs drawn from a luminosity function  as measured by   
\citet{2015AA...578A..83G}  appear to be just about sufficient to explain the observed opacity fluctuations
on large scales given the uncertainties in both the modelling and the data. 
 Note in particular that we have not taken into account in our modelling the effect of the additional photo-heating 
due to the harder spectra of the QSOs which will lead to an earlier reionization of HeII in the vicinity of the  QSOs. This should lead
to temperature fluctuations on comparable spatial scales and would lead to an amplification of the opacity fluctuations 
as low opacity regions would also be hotter and therefore have a reduced recombination rate further 
reducing the opacity.  Taking this into account should reduce the contribution of QSO to the ionizing 
emissivity  required to produce the same level of opacity fluctuations. This may be necessary 
if the contribution of QSO to the ionizing emissivity  is lower as suggested by the luminosity function
measured by \citet{2015AA...578A..83G}, what may -- as discussed in the Introduction -- well be the case. 
We have also not accounted for a possible beaming and/or (short) duty cycle of the QSOs in our modelling 
and the net effect of these on the predicted opacity fluctuations  is not obvious.

\subsection{Alternative models for the large opacity fluctuations on large scales}
\label{other_effects}

We have shown here that large amplitude UV background fluctuations 
due to a significant contribution of QSOs  may be able to explain the large reported opacity 
fluctuations. A similar conclusion has been reached by \citet{2016arXiv160706467D} 
which first became available to us during the refereeing process of this paper. However, it is not certain  that QSOs in sufficient numbers actually exist. We therefore 
briefly  discuss here also alternative suggestions. \citet{2016MNRAS.460.1328D} presented  
modelling where they reproduce the large opacity fluctuations with a hybrid DM semi-analytical
model of ionization by (faint) galaxies only. Similarly to what we have done here in the last section they assume a photoionization 
rate dependent  mean free path. They show that with  a rather low normalisation of the mean free 
path, the  mean free path in under dense regions where the emissivity drops can become very 
short. They further  show that in this way moderate spatial fluctuations in the ionizing emissivity can 
be amplified into  large scale opacity fluctuations in reasonable agreement with those reported.
Note, however, that the simulation technique of \citet{2016MNRAS.460.1328D} did not allow 
to produce realistic mock absorption spectra and that  the fluctuating Gunn-Peterson 
effect approximation was used instead.  What also remains still to be seen is how realistic the very short assumed 
mean free paths in under dense regions  are.

A further alternative explanation of the large opacity fluctuation has been 
recently proposed by \citet{2015ApJ...813L..38D}. These authors have argued
 that temperature fluctuations due to a spread of  times when hydrogen reionization 
 occurs  could explain the large opacity fluctuations on large scales 
due to the temperature dependence of the recombination rate.    Using also a hybrid-technique 
based  on high-resolution hydro-simulation combined with the excursion set formalism technique, 
 \citet{2015ApJ...813L..38D} showed that the patchy nature of reionization could lead to large scale temperature fluctuations  of sufficient 
amplitude.  They also  show that they can reproduce the reported 
$\mathrm{\tau_{eff}}$ PDF, albeit with rather extreme assumption for the
temperature of the IGM immediately after a patch is reionized. 
Reionization is however  assumed to start early and to be rather extended, somewhat at odds with recent evidence for a 
rather late reionization (\citealt{2015MNRAS.452..261C}, \citealt{2015arXiv150201589P}, \citealt{2016arXiv160503507P}) .   
\citet{2015ApJ...813L..38D} furthermore neglect the effect of helium photo-heating on the temperature (evolution) of the IGM
and assume rather high  temperatures (up to 30000K)  immediately post-reionization.

In the late stages of preparing the manuscript for this paper a preprint  by \citet{2016arXiv160503183G} appeared which found reasonable 
agreement  between  the $\tau_{\rm eff}$ PDF in   cosmological reionization simulations run as part of the CROC project. 
The simulations are similar to ours in terms of box size and resolution, but have only galaxies as ionizing sources. The
main difference in how the $\tau_{\rm eff}$ PDF was calculated appears to be that they  have averaged 
simulations with a range of mean densities of their simulation box to take into account cosmic variance. 
This approach leads to rather different redshifts where the ionized regions percolate in the individual simulations from z$\sim6.75$ to z$\sim5.5$ 
(see \citealt{2014ApJ...793...29G} and \citealt{2014ApJ...793...30G}; for comparison our simulations from paper I overlap at about z$\sim6$).
Averaging over these different reionization histories drastically widens the PDF. We think this approach is rather problematic as it 
does not allow for ionizing radiation from sources in over-dense regions to ionize under-dense regions from 
the outside and we think that this approach therefore will significantly overestimate the width of the $\tau_{\rm eff}$ PDF. 
Unfortunately, simulations of larger regions with similar resolution as we have employed will be required to get to (more) converged results. 
Unfortunately at present we were not yet able to "afford" the required dynamic range.

\subsection{Correlations with other observables}
\label{missing_quasars}

As discussed in the last section, there are now three rather different suggestions for what is primarily responsible for the 
large reported opacity fluctuations at $z\sim 5.4-5.8$  on large scales. Gratifyingly the different suggested explanations make rather 
different predictions for how the \lya opacity correlates with other observables. 

As already mentioned in the model discussed here regions of high transmitted flux 
should be correlated with the presence of QSOs near the line of sight 
(the transverse proximity  effect, see \citealt{2008MNRAS.386..359G}). The  detectability of this will depend on the QSO duty cycle 
and the angular distribution of the ionizing radiation. Better characterisation of the QSO luminosity function in particular around its
"knee" will also help to determine the contribution of QSOs to the ionizing UV background and the opacity fluctuations
at high redshift. Interesting in this regard is also the  discovery of a rather large number  of bright high-redshift galaxies with rather hard spectra 
(see \citealt{2014ApJ...792...76B}, \citealt{2014MNRAS.440.2810B}, \citealt{2015MNRAS.454.1393S} and \citealt{2016arXiv160601304S}).

The model of \citet{2016MNRAS.460.1328D} predicts  a very strong anti-correlation of the ionizing emissivity of galaxies with the long absorption troughs falling 
into large regions below the average ionizing emissivity. As already discussed it  also predicts a rather rapid decrease 
of the mean free path at $z>5.5$ in particular in regions underdense in ionizing sources. 

The observational signatures 
of the temperature fluctuation model of \citet{2015ApJ...813L..38D} are less clear,  but there should be a correlation rather than an anti-correlation between 
density field (smoothed on) on large scales and $\mathrm{\tau_{eff}}$ as the overdense regions should be ionized earlier  
and should have more time to cool down to lower temperature leading to larger $\mathrm{\tau_{eff}}$. As already mentioned  this model predicts a 
rather early start of reionization.

There is of course the possibility that more than one  of the effects 
discussed above contribute significantly to the opacity fluctuations on large scales.
As already pointed out, the models of \citet{2015ApJ...813L..38D} and \citet{2016MNRAS.460.1328D} predict opposite observational signatures. 
Therefore we can imagine that if the two effects play an equivalent role, they could potentially cancel each other and might thereby not be able to explain 
the scatter in the $\mathrm{\tau_{eff}}$ PDF.

Our model with AGNs does yet not take the temperature effects invoked by \citet{2015ApJ...813L..38D} into account. We expect that the proximity zones
of quasars will still be high transmissivity regions when this is included. While hydrogen might reionize early there, these regions are likely 
nevertheless hot due to HeII getting ionized there early as well.  
It is difficult to guess if the global shape of the effective \lya opacity PDF would change or not.
More sophisticated simulations including both galaxies and AGNs with a proper treatment of the temperature evolution will be needed to answer this question.

The best way to test the additional effect of the fluctuating mean free path on large scales invoked by \citet{2016MNRAS.460.1328D} 
would be to run much larger simulations of reionization by galaxies to get a fair sampling of the voids 
and the related spatial distribution of the mean free path. 
For the AGNs, the model presented in this work already accounts in a simplified manner for the expected spatial fluctuation of the mean 
free path. However a more accurate treatment of the modulation of the mean free path in a scenario where both AGNs and galaxies act as ionizing sources
would require full multi-frequency radiative transfer simulations in very large volumes, which is  very difficult/not yet possible with current facilities at the desired resolution.

One interesting idea to disentangle the different scenarios would be to have a closer look to the line of sight of the quasar ULAS J0148+0600 
that shows a very long ($110 h^{-1}$ comoving Mpc)  high optical depth Gunn Peterson trough.
The goal would be to measure the over/under-density of galaxies in a cylinder around the trough to 
understand if high opacity regions are more prone to exhibit an excess or dearth of galaxies around them.
The results could potentially give some constraints in regards to the model of \citet{2016MNRAS.460.1328D} and the one of \citet{2015ApJ...813L..38D}.


\section{Conclusions and Outlook}
\label{prospects}

We have combined here  high-resolution  full radiative transfer simulations 
with the large volume  Millennium simulation to model large scale opacity fluctuations 
due to a significant contribution of QSOs  to the UV background at$z\ga 5$.

Our main results are as follows. 

\begin{itemize} 
\item{We can reproduce the reported broad distribution of  the \lya opacity on scales $\ge 50 h^{-1}$ comoving Mpc  
         with a  contribution $\ga$50\%  of QSOs to the ionizing emissivity.}
\item{The ionizing emissivity of QSO required to reproduce the observed opacity depends rather sensitively 
         on the assumed mean free path and its dependence on the  local ionizing UV flux and over-density.} 
\item{For assumptions    for the mean free path and its dependence suggested by our simulations the required ionizing 
         emissivity  is similar to that predicted 
         by the recent determination of the QSO luminosity function at this redshift by \citet{2015AA...578A..83G}.}

\item{Our simulations neglect the increased temperatures around QSOs 
         (primarily due to the reionization of HeII)  which will lead to correlated temperature  and opacity fluctuations on large scales.
         Accounting  for this should  further reduce the ionizing emissivity due to QSOs required to explain the 
          large opacity fluctuations on large scales.} 
\item{Our simulations also reproduce very long ($110 h^{-1}$ comoving Mpc)  high optical depth Gunn Peterson troughs 
         like the one reported for  ULAS J0148+0600, albeit rather rarely (in $\sim$ 7\% of spectra).}   
\item{The model predicts  a strong correlation of low  \lya opacity with the presence of QSOs close to the line-of-sight.
         This  differs strongly  from the predictions of alternative models that predict a strong correlation or anti-correlation 
          of \lya opacity with (over-) density on large scales. The strength of the correlation should depend on the duty cycle as well as the 
          the possible beaming of the QSOs.}

\end{itemize}          
          
Note, however, that the simulations presented here still suffer from their limited   dynamic range. In particular 
our  high-resolution hydro/RT simulations while (marginally)  resolving the \lya opacity (as well as the sinks of ionizing radiation) do 
not have a large enough  simulation volume to comfortably capture the mean free-path after percolation of the HII regions due to galaxies.  
Furthermore, the modelling of the QSO contribution to the ionizing UV background could only  be done  in post-processing 
with a rather simple model of the effect of QSOs on the  mean free path. 
The modelling of  alternative explanations for the reported large scale opacity fluctuations, large scale emissivity fluctuations of the galaxy ionizing 
emissivity and temperature fluctuations due to 
"patchy" reionization suffers similar limitations.  

Pushing our numerical simulations to higher dynamic range i.e. 
larger volumes at comparable resolution and more sophisticated (full RT) modelling of the QSO contribution, including light 
travel effects for the QSO, exploring the effect of varying the angular distribution of their radiation and 
improved  modelling of  the photo-heating of hydrogen and in particular hydrogen with multi-frequency RT simulations
will be all important  for improving the robustness of the predictions. This should, however,  be well worth the effort, 
as it will allow to infer much of what happened during the epoch of reionization from the fossil record present in \lya forest data 
in the post-reionization epoch.

\section*{Acknowledgments}

We thank Dominique Aubert for letting us use his code ATON and for helpful discussions. 
We thank George Becker for helpful communication
regarding his measurements of the flux PDF.
We further thank James Bolton and Xiahoui Fan for helpful comments and suggestions.
This work was supported by the ERC Advanced Grant 320596
``The Emergence of Structure during the epoch of Reionization". 
This work was performed using the COSMOS and Darwin Supercomputers of the University of Cambridge High Performance Computing Service (http://www.hpc.cam.ac.uk/), 
provided by Dell Inc. using Strategic Research Infrastructure Funding from the Higher Education Funding Council for England and funding from the Science and Technology Facilities Council.
This work furthermore  used the Wilkes GPU cluster at the University of Cambridge High Performance Computing Service (http://www.hpc.cam.ac.uk/), provided by Dell Inc., 
NVIDIA and Mellanox, and part funded by STFC with industrial sponsorship from Rolls Royce and Mitsubishi Heavy Industries.
Some of the hydrodynamical simulations used in this work were performed with
supercomputer time awarded by the Partnership for Advanced Computing in Europe
(PRACE)  8th Call. We acknowledge PRACE for enabling access to the Curie supercomputer,
based in France at the Tr\`{e}s Grand Centre de Calcul (TGCC).
EP acknowledges support by the Kavli foundation. 
The research of  JC and MH was supported by the Munich Institute for Astro- and Particle Physics (MIAPP) 
of the DFG cluster of excellence ``Origin and Structure of the Universe''.

\bibliographystyle{mnras}
\bibliography{biblio}

\begin{thebibliography}{}
\makeatletter
\relax
\def\mn@urlcharsother{\let\do\@makeother \do\$\do\&\do\#\do\^\do\_\do\%\do\~}
\def\mn@doi{\begingroup\mn@urlcharsother \@ifnextchar [ {\mn@doi@}
  {\mn@doi@[]}}
\def\mn@doi@[#1]#2{\def\@tempa{#1}\ifx\@tempa\@empty \href
  {http://dx.doi.org/#2} {doi:#2}\else \href {http://dx.doi.org/#2} {#1}\fi
  \endgroup}
\def\mn@eprint#1#2{\mn@eprint@#1:#2::\@nil}
\def\mn@eprint@arXiv#1{\href {http://arxiv.org/abs/#1} {{\tt arXiv:#1}}}
\def\mn@eprint@dblp#1{\href {http://dblp.uni-trier.de/rec/bibtex/#1.xml}
  {dblp:#1}}
\def\mn@eprint@#1:#2:#3:#4\@nil{\def\@tempa {#1}\def\@tempb {#2}\def\@tempc
  {#3}\ifx \@tempc \@empty \let \@tempc \@tempb \let \@tempb \@tempa \fi \ifx
  \@tempb \@empty \def\@tempb {arXiv}\fi \@ifundefined
  {mn@eprint@\@tempb}{\@tempb:\@tempc}{\expandafter \expandafter \csname
  mn@eprint@\@tempb\endcsname \expandafter{\@tempc}}}

\bibitem[\protect\citeauthoryear{{Aubert} \& {Teyssier}}{{Aubert} \&
  {Teyssier}}{2008}]{2008MNRAS.387..295A}
{Aubert} D.,  {Teyssier} R.,  2008, \mn@doi [\mnras]
  {10.1111/j.1365-2966.2008.13223.x}, \href
  {http://cdsads.u-strasbg.fr/abs/2008MNRAS.387..295A} {387, 295}

\bibitem[\protect\citeauthoryear{{Aubert}, {Deparis}  \& {Ocvirk}}{{Aubert}
  et~al.}{2015}]{2015MNRAS.454.1012A}
{Aubert} D.,  {Deparis} N.,   {Ocvirk} P.,  2015, \mn@doi [\mnras]
  {10.1093/mnras/stv1896}, \href
  {http://adsabs.harvard.edu/abs/2015MNRAS.454.1012A} {454, 1012}

\bibitem[\protect\citeauthoryear{{Baek}, {Di Matteo}, {Semelin}, {Combes}  \&
  {Revaz}}{{Baek} et~al.}{2009}]{2009A&A...495..389B}
{Baek} S.,  {Di Matteo} P.,  {Semelin} B.,  {Combes} F.,   {Revaz} Y.,  2009,
  \mn@doi [\aap] {10.1051/0004-6361:200810757}, \href
  {http://cdsads.u-strasbg.fr/abs/2009A\%26A...495..389B} {495, 389}

\bibitem[\protect\citeauthoryear{{Becker} \& {Bolton}}{{Becker} \&
  {Bolton}}{2013}]{2013MNRAS.436.1023B}
{Becker} G.~D.,  {Bolton} J.~S.,  2013, \mn@doi [\mnras]
  {10.1093/mnras/stt1610}, \href
  {http://adsabs.harvard.edu/abs/2013MNRAS.436.1023B} {436, 1023}

\bibitem[\protect\citeauthoryear{{Becker}, {Bolton}, {Haehnelt}  \&
  {Sargent}}{{Becker} et~al.}{2011}]{2011MNRAS.410.1096B}
{Becker} G.~D.,  {Bolton} J.~S.,  {Haehnelt} M.~G.,   {Sargent} W.~L.~W.,
  2011, \mn@doi [\mnras] {10.1111/j.1365-2966.2010.17507.x}, \href
  {http://adsabs.harvard.edu/abs/2011MNRAS.410.1096B} {410, 1096}

\bibitem[\protect\citeauthoryear{{Becker}, {Hewett}, {Worseck}  \&
  {Prochaska}}{{Becker} et~al.}{2013}]{2013MNRAS.430.2067B}
{Becker} G.~D.,  {Hewett} P.~C.,  {Worseck} G.,   {Prochaska} J.~X.,  2013,
  \mn@doi [\mnras] {10.1093/mnras/stt031}, \href
  {http://adsabs.harvard.edu/abs/2013MNRAS.430.2067B} {430, 2067}

\bibitem[\protect\citeauthoryear{{Becker}, {Bolton}  \& {Lidz}}{{Becker}
  et~al.}{2015a}]{2015PASA...32...45B}
{Becker} G.~D.,  {Bolton} J.~S.,   {Lidz} A.,  2015a, \mn@doi [\pasa]
  {10.1017/pasa.2015.45}, \href
  {http://adsabs.harvard.edu/abs/2015PASA...32...45B} {32, e045}

\bibitem[\protect\citeauthoryear{{Becker}, {Bolton}, {Madau}, {Pettini},
  {Ryan-Weber}  \& {Venemans}}{{Becker} et~al.}{2015b}]{2015MNRAS.447.3402B}
{Becker} G.~D.,  {Bolton} J.~S.,  {Madau} P.,  {Pettini} M.,  {Ryan-Weber}
  E.~V.,   {Venemans} B.~P.,  2015b, \mn@doi [\mnras] {10.1093/mnras/stu2646},
  \href {http://adsabs.harvard.edu/abs/2015MNRAS.447.3402B} {447, 3402}

\bibitem[\protect\citeauthoryear{{Boera}, {Murphy}, {Becker}  \&
  {Bolton}}{{Boera} et~al.}{2014}]{2014MNRAS.441.1916B}
{Boera} E.,  {Murphy} M.~T.,  {Becker} G.~D.,   {Bolton} J.~S.,  2014, \mn@doi
  [\mnras] {10.1093/mnras/stu660}, \href
  {http://adsabs.harvard.edu/abs/2014MNRAS.441.1916B} {441, 1916}

\bibitem[\protect\citeauthoryear{{Bolton} \& {Haehnelt}}{{Bolton} \&
  {Haehnelt}}{2007}]{2007MNRAS.382..325B}
{Bolton} J.~S.,  {Haehnelt} M.~G.,  2007, \mn@doi [\mnras]
  {10.1111/j.1365-2966.2007.12372.x}, \href
  {http://adsabs.harvard.edu/abs/2007MNRAS.382..325B} {382, 325}

\bibitem[\protect\citeauthoryear{{Bolton} \& {Haehnelt}}{{Bolton} \&
  {Haehnelt}}{2013}]{2013MNRAS.429.1695B}
{Bolton} J.~S.,  {Haehnelt} M.~G.,  2013, \mn@doi [\mnras]
  {10.1093/mnras/sts455}, \href
  {http://adsabs.harvard.edu/abs/2013MNRAS.429.1695B} {429, 1695}

\bibitem[\protect\citeauthoryear{{Bolton} \& {Viel}}{{Bolton} \&
  {Viel}}{2011}]{2011MNRAS.414..241B}
{Bolton} J.~S.,  {Viel} M.,  2011, \mn@doi [\mnras]
  {10.1111/j.1365-2966.2011.18384.x}, \href
  {http://adsabs.harvard.edu/abs/2011MNRAS.414..241B} {414, 241}

\bibitem[\protect\citeauthoryear{{Bolton}, {Puchwein}, {Sijacki}, {Haehnelt},
  {Kim}, {Meiksin}, {Regan}  \& {Viel}}{{Bolton}
  et~al.}{2016}]{2016arXiv160503462B}
{Bolton} J.~S.,  {Puchwein} E.,  {Sijacki} D.,  {Haehnelt} M.~G.,  {Kim} T.-S.,
   {Meiksin} A.,  {Regan} J.~A.,   {Viel} M.,  2016, preprint, \href
  {http://adsabs.harvard.edu/abs/2016arXiv160503462B} {} (\mn@eprint {arXiv}
  {1605.03462})

\bibitem[\protect\citeauthoryear{{Borthakur}, {Heckman}, {Leitherer}  \&
  {Overzier}}{{Borthakur} et~al.}{2014}]{2014Sci...346..216B}
{Borthakur} S.,  {Heckman} T.~M.,  {Leitherer} C.,   {Overzier} R.~A.,  2014,
  \mn@doi [Science] {10.1126/science.1254214}, \href
  {http://adsabs.harvard.edu/abs/2014Sci...346..216B} {346, 216}

\bibitem[\protect\citeauthoryear{{Bouwens}, {Illingworth}, {Franx}  \&
  {Ford}}{{Bouwens} et~al.}{2007}]{2007ApJ...670..928B}
{Bouwens} R.~J.,  {Illingworth} G.~D.,  {Franx} M.,   {Ford} H.,  2007, \mn@doi
  [\apj] {10.1086/521811}, \href
  {http://adsabs.harvard.edu/abs/2007ApJ...670..928B} {670, 928}

\bibitem[\protect\citeauthoryear{{Bouwens} et~al.,}{{Bouwens}
  et~al.}{2015}]{2015ApJ...803...34B}
{Bouwens} R.~J.,  et~al., 2015, \mn@doi [\apj] {10.1088/0004-637X/803/1/34},
  \href {http://adsabs.harvard.edu/abs/2015ApJ...803...34B} {803, 34}

\bibitem[\protect\citeauthoryear{{Bowler} et~al.,}{{Bowler}
  et~al.}{2014}]{2014MNRAS.440.2810B}
{Bowler} R.~A.~A.,  et~al., 2014, \mn@doi [\mnras] {10.1093/mnras/stu449},
  \href {http://adsabs.harvard.edu/abs/2014MNRAS.440.2810B} {440, 2810}

\bibitem[\protect\citeauthoryear{{Bradley} et~al.,}{{Bradley}
  et~al.}{2014}]{2014ApJ...792...76B}
{Bradley} L.~D.,  et~al., 2014, \mn@doi [\apj] {10.1088/0004-637X/792/1/76},
  \href {http://adsabs.harvard.edu/abs/2014ApJ...792...76B} {792, 76}

\bibitem[\protect\citeauthoryear{{Calverley}, {Becker}, {Haehnelt}  \&
  {Bolton}}{{Calverley} et~al.}{2011}]{2011MNRAS.412.2543C}
{Calverley} A.~P.,  {Becker} G.~D.,  {Haehnelt} M.~G.,   {Bolton} J.~S.,  2011,
  \mn@doi [\mnras] {10.1111/j.1365-2966.2010.18072.x}, \href
  {http://adsabs.harvard.edu/abs/2011MNRAS.412.2543C} {412, 2543}

\bibitem[\protect\citeauthoryear{{Chardin}, {Aubert}  \& {Ocvirk}}{{Chardin}
  et~al.}{2012}]{2012A&A...548A...9C}
{Chardin} J.,  {Aubert} D.,   {Ocvirk} P.,  2012, \mn@doi [\aap]
  {10.1051/0004-6361/201219992}, \href
  {http://adsabs.harvard.edu/abs/2012A\%26A...548A...9C} {548, A9}

\bibitem[\protect\citeauthoryear{{Chardin}, {Aubert}  \& {Ocvirk}}{{Chardin}
  et~al.}{2014}]{2014A&A...568A..52C}
{Chardin} J.,  {Aubert} D.,   {Ocvirk} P.,  2014, \mn@doi [\aap]
  {10.1051/0004-6361/201322355}, \href
  {http://adsabs.harvard.edu/abs/2014A%26A...568A..52C} {568, A52}

\bibitem[\protect\citeauthoryear{{Chardin}, {Haehnelt}, {Aubert}  \&
  {Puchwein}}{{Chardin} et~al.}{2015}]{2015MNRAS.453.2943C}
{Chardin} J.,  {Haehnelt} M.~G.,  {Aubert} D.,   {Puchwein} E.,  2015, \mn@doi
  [\mnras] {10.1093/mnras/stv1786}, \href
  {http://adsabs.harvard.edu/abs/2015MNRAS.453.2943C} {453, 2943}

\bibitem[\protect\citeauthoryear{{Choudhury}, {Puchwein}, {Haehnelt}  \&
  {Bolton}}{{Choudhury} et~al.}{2015}]{2015MNRAS.452..261C}
{Choudhury} T.~R.,  {Puchwein} E.,  {Haehnelt} M.~G.,   {Bolton} J.~S.,  2015,
  \mn@doi [\mnras] {10.1093/mnras/stv1250}, \href
  {http://adsabs.harvard.edu/abs/2015MNRAS.452..261C} {452, 261}

\bibitem[\protect\citeauthoryear{{Ciardi} \& {Ferrara}}{{Ciardi} \&
  {Ferrara}}{2005}]{2005SSRv..116..625C}
{Ciardi} B.,  {Ferrara} A.,  2005, \mn@doi [\ssr] {10.1007/s11214-005-3592-0},
  \href {http://adsabs.harvard.edu/abs/2005SSRv..116..625C} {116, 625}

\bibitem[\protect\citeauthoryear{{D'Aloisio}, {McQuinn}  \& {Trac}}{{D'Aloisio}
  et~al.}{2015}]{2015ApJ...813L..38D}
{D'Aloisio} A.,  {McQuinn} M.,   {Trac} H.,  2015, \mn@doi [\apjl]
  {10.1088/2041-8205/813/2/L38}, \href
  {http://adsabs.harvard.edu/abs/2015ApJ...813L..38D} {813, L38}

\bibitem[\protect\citeauthoryear{{D'Aloisio}, {Upton Sanderbeck}, {McQuinn},
  {Trac}  \& {Shapiro}}{{D'Aloisio} et~al.}{2016}]{2016arXiv160706467D}
{D'Aloisio} A.,  {Upton Sanderbeck} P.~R.,  {McQuinn} M.,  {Trac} H.,
  {Shapiro} P.~R.,  2016, preprint, \href
  {http://adsabs.harvard.edu/abs/2016arXiv160706467D} {} (\mn@eprint {arXiv}
  {1607.06467})

\bibitem[\protect\citeauthoryear{{Davies} \& {Furlanetto}}{{Davies} \&
  {Furlanetto}}{2016}]{2016MNRAS.460.1328D}
{Davies} F.~B.,  {Furlanetto} S.~R.,  2016, \mn@doi [\mnras]
  {10.1093/mnras/stw931}, \href
  {http://adsabs.harvard.edu/abs/2016MNRAS.460.1328D} {460, 1328}

\bibitem[\protect\citeauthoryear{{Dijkstra}, {Wyithe}, {Haiman}, {Mesinger}  \&
  {Pentericci}}{{Dijkstra} et~al.}{2014}]{2014MNRAS.440.3309D}
{Dijkstra} M.,  {Wyithe} S.,  {Haiman} Z.,  {Mesinger} A.,   {Pentericci} L.,
  2014, \mn@doi [\mnras] {10.1093/mnras/stu531}, \href
  {http://adsabs.harvard.edu/abs/2014MNRAS.440.3309D} {440, 3309}

\bibitem[\protect\citeauthoryear{{Fan} et~al.,}{{Fan}
  et~al.}{2006}]{2006AJ....132..117F}
{Fan} X.,  et~al., 2006, \mn@doi [\aj] {10.1086/504836}, \href
  {http://cdsads.u-strasbg.fr/abs/2006AJ....132..117F} {132, 117}

\bibitem[\protect\citeauthoryear{{Finkelstein} et~al.,}{{Finkelstein}
  et~al.}{2015}]{2015ApJ...810...71F}
{Finkelstein} S.~L.,  et~al., 2015, \mn@doi [\apj]
  {10.1088/0004-637X/810/1/71}, \href
  {http://adsabs.harvard.edu/abs/2015ApJ...810...71F} {810, 71}

\bibitem[\protect\citeauthoryear{{Fotopoulou} et~al.,}{{Fotopoulou}
  et~al.}{2016}]{2016arXiv160303240F}
{Fotopoulou} S.,  et~al., 2016, preprint, \href
  {http://adsabs.harvard.edu/abs/2016arXiv160303240F} {} (\mn@eprint {arXiv}
  {1603.03240})

\bibitem[\protect\citeauthoryear{{Gallerani}, {Ferrara}, {Fan}  \&
  {Choudhury}}{{Gallerani} et~al.}{2008}]{2008MNRAS.386..359G}
{Gallerani} S.,  {Ferrara} A.,  {Fan} X.,   {Choudhury} T.~R.,  2008, \mn@doi
  [\mnras] {10.1111/j.1365-2966.2008.13029.x}, \href
  {http://adsabs.harvard.edu/abs/2008MNRAS.386..359G} {386, 359}

\bibitem[\protect\citeauthoryear{{Georgakakis} et~al.,}{{Georgakakis}
  et~al.}{2015}]{2015MNRAS.453.1946G}
{Georgakakis} A.,  et~al., 2015, \mn@doi [\mnras] {10.1093/mnras/stv1703},
  \href {http://adsabs.harvard.edu/abs/2015MNRAS.453.1946G} {453, 1946}

\bibitem[\protect\citeauthoryear{{Giallongo} et~al.,}{{Giallongo}
  et~al.}{2015}]{2015AA...578A..83G}
{Giallongo} E.,  et~al., 2015, \mn@doi [\aap] {10.1051/0004-6361/201425334},
  \href {http://adsabs.harvard.edu/abs/2015A%26A...578A..83G} {578, A83}

\bibitem[\protect\citeauthoryear{{Gnedin}}{{Gnedin}}{2014}]{2014ApJ...793...29G}
{Gnedin} N.~Y.,  2014, \mn@doi [\apj] {10.1088/0004-637X/793/1/29}, \href
  {http://adsabs.harvard.edu/abs/2014ApJ...793...29G} {793, 29}

\bibitem[\protect\citeauthoryear{{Gnedin} \& {Kaurov}}{{Gnedin} \&
  {Kaurov}}{2014}]{2014ApJ...793...30G}
{Gnedin} N.~Y.,  {Kaurov} A.~A.,  2014, \mn@doi [\apj]
  {10.1088/0004-637X/793/1/30}, \href
  {http://adsabs.harvard.edu/abs/2014ApJ...793...30G} {793, 30}

\bibitem[\protect\citeauthoryear{{Gnedin}, {Becker}  \& {Fan}}{{Gnedin}
  et~al.}{2016}]{2016arXiv160503183G}
{Gnedin} N.~Y.,  {Becker} G.~D.,   {Fan} X.,  2016, preprint, \href
  {http://adsabs.harvard.edu/abs/2016arXiv160503183G} {} (\mn@eprint {arXiv}
  {1605.03183})

\bibitem[\protect\citeauthoryear{{Grazian} et~al.,}{{Grazian}
  et~al.}{2016}]{2016A&A...585A..48G}
{Grazian} A.,  et~al., 2016, \mn@doi [\aap] {10.1051/0004-6361/201526396},
  \href {http://adsabs.harvard.edu/abs/2016A%26A...585A..48G} {585, A48}

\bibitem[\protect\citeauthoryear{{Haardt} \& {Madau}}{{Haardt} \&
  {Madau}}{2012}]{2012ApJ...746..125H}
{Haardt} F.,  {Madau} P.,  2012, \mn@doi [\apj] {10.1088/0004-637X/746/2/125},
  \href {http://adsabs.harvard.edu/abs/2012ApJ...746..125H} {746, 125}

\bibitem[\protect\citeauthoryear{{Haardt} \& {Salvaterra}}{{Haardt} \&
  {Salvaterra}}{2015}]{2015A&A...575L..16H}
{Haardt} F.,  {Salvaterra} R.,  2015, \mn@doi [\aap]
  {10.1051/0004-6361/201525627}, \href
  {http://cdsads.u-strasbg.fr/abs/2015A%26A...575L..16H} {575, L16}

\bibitem[\protect\citeauthoryear{{Hopkins}, {Richards}  \&
  {Hernquist}}{{Hopkins} et~al.}{2007}]{2007ApJ...654..731H}
{Hopkins} P.~F.,  {Richards} G.~T.,   {Hernquist} L.,  2007, \mn@doi [\apj]
  {10.1086/509629}, \href {http://adsabs.harvard.edu/abs/2007ApJ...654..731H}
  {654, 731}

\bibitem[\protect\citeauthoryear{{Hui} \& {Gnedin}}{{Hui} \&
  {Gnedin}}{1997}]{1997MNRAS.292...27H}
{Hui} L.,  {Gnedin} N.~Y.,  1997, \mnras, \href
  {http://adsabs.harvard.edu/abs/1997MNRAS.292...27H} {292, 27}

\bibitem[\protect\citeauthoryear{{Iliev}, {Mellema}, {Pen}, {Merz}, {Shapiro}
  \& {Alvarez}}{{Iliev} et~al.}{2006}]{2006MNRAS.369.1625I}
{Iliev} I.~T.,  {Mellema} G.,  {Pen} U.,  {Merz} H.,  {Shapiro} P.~R.,
  {Alvarez} M.~A.,  2006, \mn@doi [\mnras] {10.1111/j.1365-2966.2006.10502.x},
  \href {http://cdsads.u-strasbg.fr/abs/2006MNRAS.369.1625I} {369, 1625}

\bibitem[\protect\citeauthoryear{{Jiang} et~al.,}{{Jiang}
  et~al.}{2009}]{2009AJ....138..305J}
{Jiang} L.,  et~al., 2009, \mn@doi [\aj] {10.1088/0004-6256/138/1/305}, \href
  {http://adsabs.harvard.edu/abs/2009AJ....138..305J} {138, 305}

\bibitem[\protect\citeauthoryear{{Kimm} \& {Cen}}{{Kimm} \&
  {Cen}}{2014}]{2014ApJ...788..121K}
{Kimm} T.,  {Cen} R.,  2014, \mn@doi [\apj] {10.1088/0004-637X/788/2/121},
  \href {http://adsabs.harvard.edu/abs/2014ApJ...788..121K} {788, 121}

\bibitem[\protect\citeauthoryear{{Kimm}, {Katz}, {Haehnelt}, {Rosdahl},
  {Devriendt}  \& {Slyz}}{{Kimm} et~al.}{2016}]{2016arXiv160804762K}
{Kimm} T.,  {Katz} H.,  {Haehnelt} M.,  {Rosdahl} J.,  {Devriendt} J.,   {Slyz}
  A.,  2016, preprint, \href
  {http://adsabs.harvard.edu/abs/2016arXiv160804762K} {} (\mn@eprint {arXiv}
  {1608.04762})

\bibitem[\protect\citeauthoryear{{Kulkarni}, {Hennawi}, {O{\~n}orbe}, {Rorai}
  \& {Springel}}{{Kulkarni} et~al.}{2015}]{2015ApJ...812...30K}
{Kulkarni} G.,  {Hennawi} J.~F.,  {O{\~n}orbe} J.,  {Rorai} A.,   {Springel}
  V.,  2015, \mn@doi [\apj] {10.1088/0004-637X/812/1/30}, \href
  {http://adsabs.harvard.edu/abs/2015ApJ...812...30K} {812, 30}

\bibitem[\protect\citeauthoryear{{Leethochawalit}, {Jones}, {Ellis}, {Stark}
  \& {Zitrin}}{{Leethochawalit} et~al.}{2016}]{2016arXiv160605309L}
{Leethochawalit} N.,  {Jones} T.~A.,  {Ellis} R.~S.,  {Stark} D.~P.,   {Zitrin}
  A.,  2016, preprint, \href
  {http://adsabs.harvard.edu/abs/2016arXiv160605309L} {} (\mn@eprint {arXiv}
  {1606.05309})

\bibitem[\protect\citeauthoryear{{Leitherer}, {Hernandez}, {Lee}  \&
  {Oey}}{{Leitherer} et~al.}{2016}]{2016ApJ...823...64L}
{Leitherer} C.,  {Hernandez} S.,  {Lee} J.~C.,   {Oey} M.~S.,  2016, \mn@doi
  [\apj] {10.3847/0004-637X/823/1/64}, \href
  {http://adsabs.harvard.edu/abs/2016ApJ...823...64L} {823, 64}

\bibitem[\protect\citeauthoryear{{Lidz}, {McQuinn}, {Zaldarriaga}, {Hernquist}
  \& {Dutta}}{{Lidz} et~al.}{2007}]{2007ApJ...670...39L}
{Lidz} A.,  {McQuinn} M.,  {Zaldarriaga} M.,  {Hernquist} L.,   {Dutta} S.,
  2007, \mn@doi [\apj] {10.1086/521974}, \href
  {http://adsabs.harvard.edu/abs/2007ApJ...670...39L} {670, 39}

\bibitem[\protect\citeauthoryear{{Ma}, {Kasen}, {Hopkins},
  {Faucher-Gigu{\`e}re}, {Quataert}, {Kere{\v s}}  \& {Murray}}{{Ma}
  et~al.}{2015}]{2015MNRAS.453..960M}
{Ma} X.,  {Kasen} D.,  {Hopkins} P.~F.,  {Faucher-Gigu{\`e}re} C.-A.,
  {Quataert} E.,  {Kere{\v s}} D.,   {Murray} N.,  2015, \mn@doi [\mnras]
  {10.1093/mnras/stv1679}, \href
  {http://adsabs.harvard.edu/abs/2015MNRAS.453..960M} {453, 960}

\bibitem[\protect\citeauthoryear{{Madau} \& {Haardt}}{{Madau} \&
  {Haardt}}{2015}]{2015ApJ...813L...8M}
{Madau} P.,  {Haardt} F.,  2015, \mn@doi [\apjl] {10.1088/2041-8205/813/1/L8},
  \href {http://adsabs.harvard.edu/abs/2015ApJ...813L...8M} {813, L8}

\bibitem[\protect\citeauthoryear{{Madau}, {Haardt}  \& {Rees}}{{Madau}
  et~al.}{1999}]{1999ApJ...514..648M}
{Madau} P.,  {Haardt} F.,   {Rees} M.~J.,  1999, \mn@doi [\apj]
  {10.1086/306975}, \href {http://adsabs.harvard.edu/abs/1999ApJ...514..648M}
  {514, 648}

\bibitem[\protect\citeauthoryear{{Matsuoka} et~al.,}{{Matsuoka}
  et~al.}{2016}]{2016arXiv160302281M}
{Matsuoka} Y.,  et~al., 2016, preprint, \href
  {http://adsabs.harvard.edu/abs/2016arXiv160302281M} {} (\mn@eprint {arXiv}
  {1603.02281})

\bibitem[\protect\citeauthoryear{{McQuinn}}{{McQuinn}}{2015}]{2015arXiv151200086M}
{McQuinn} M.,  2015, preprint, \href
  {http://adsabs.harvard.edu/abs/2015arXiv151200086M} {} (\mn@eprint {arXiv}
  {1512.00086})

\bibitem[\protect\citeauthoryear{{McQuinn}, {Oh}  \&
  {Faucher-Gigu{\`e}re}}{{McQuinn} et~al.}{2011}]{2011ApJ...743...82M}
{McQuinn} M.,  {Oh} S.~P.,   {Faucher-Gigu{\`e}re} C.-A.,  2011, \mn@doi [\apj]
  {10.1088/0004-637X/743/1/82}, \href
  {http://adsabs.harvard.edu/abs/2011ApJ...743...82M} {743, 82}

\bibitem[\protect\citeauthoryear{{Meiksin}}{{Meiksin}}{2009}]{2009RvMP...81.1405M}
{Meiksin} A.~A.,  2009, \mn@doi [Reviews of Modern Physics]
  {10.1103/RevModPhys.81.1405}, \href
  {http://adsabs.harvard.edu/abs/2009RvMP...81.1405M} {81, 1405}

\bibitem[\protect\citeauthoryear{{Miralda-Escud{\'e}}, {Haehnelt}  \&
  {Rees}}{{Miralda-Escud{\'e}} et~al.}{2000}]{2000ApJ...530....1M}
{Miralda-Escud{\'e}} J.,  {Haehnelt} M.,   {Rees} M.~J.,  2000, \mn@doi [\apj]
  {10.1086/308330}, \href {http://cdsads.u-strasbg.fr/abs/2000ApJ...530....1M}
  {530, 1}

\bibitem[\protect\citeauthoryear{{Mostardi}, {Shapley}, {Steidel}, {Trainor},
  {Reddy}  \& {Siana}}{{Mostardi} et~al.}{2015}]{2015ApJ...810..107M}
{Mostardi} R.~E.,  {Shapley} A.~E.,  {Steidel} C.~C.,  {Trainor} R.~F.,
  {Reddy} N.~A.,   {Siana} B.,  2015, \mn@doi [\apj]
  {10.1088/0004-637X/810/2/107}, \href
  {http://adsabs.harvard.edu/abs/2015ApJ...810..107M} {810, 107}

\bibitem[\protect\citeauthoryear{{Ocvirk} et~al.,}{{Ocvirk}
  et~al.}{2015}]{2015arXiv151100011O}
{Ocvirk} P.,  et~al., 2015, preprint, \href
  {http://adsabs.harvard.edu/abs/2015arXiv151100011O} {} (\mn@eprint {arXiv}
  {1511.00011})

\bibitem[\protect\citeauthoryear{{Pawlik}, {Schaye}  \& {Dalla
  Vecchia}}{{Pawlik} et~al.}{2015}]{2015MNRAS.451.1586P}
{Pawlik} A.~H.,  {Schaye} J.,   {Dalla Vecchia} C.,  2015, \mn@doi [\mnras]
  {10.1093/mnras/stv976}, \href
  {http://adsabs.harvard.edu/abs/2015MNRAS.451.1586P} {451, 1586}

\bibitem[\protect\citeauthoryear{{Planck Collaboration} et~al.,}{{Planck
  Collaboration} et~al.}{2015}]{2015arXiv150201589P}
{Planck Collaboration} et~al., 2015, ArXiv e-prints 1502.01589, \href
  {http://adsabs.harvard.edu/abs/2015arXiv150201589P} {}

\bibitem[\protect\citeauthoryear{{Planck Collaboration} et~al.,}{{Planck
  Collaboration} et~al.}{2016}]{2016arXiv160503507P}
{Planck Collaboration} et~al., 2016, preprint, \href
  {http://adsabs.harvard.edu/abs/2016arXiv160503507P} {} (\mn@eprint {arXiv}
  {1605.03507})

\bibitem[\protect\citeauthoryear{{Puchwein} \& {Springel}}{{Puchwein} \&
  {Springel}}{2013}]{2013MNRAS.428.2966P}
{Puchwein} E.,  {Springel} V.,  2013, \mn@doi [\mnras] {10.1093/mnras/sts243},
  \href {http://adsabs.harvard.edu/abs/2013MNRAS.428.2966P} {428, 2966}

\bibitem[\protect\citeauthoryear{{Rahmati}, {Pawlik}, {Rai{\v c}evic}  \&
  {Schaye}}{{Rahmati} et~al.}{2013}]{2013MNRAS.430.2427R}
{Rahmati} A.,  {Pawlik} A.~H.,  {Rai{\v c}evic} M.,   {Schaye} J.,  2013,
  \mn@doi [\mnras] {10.1093/mnras/stt066}, \href
  {http://adsabs.harvard.edu/abs/2013MNRAS.430.2427R} {430, 2427}

\bibitem[\protect\citeauthoryear{{Rahmati}, {Schaye}, {Bower}, {Crain},
  {Furlong}, {Schaller}  \& {Theuns}}{{Rahmati}
  et~al.}{2015}]{2015MNRAS.452.2034R}
{Rahmati} A.,  {Schaye} J.,  {Bower} R.~G.,  {Crain} R.~A.,  {Furlong} M.,
  {Schaller} M.,   {Theuns} T.,  2015, \mn@doi [\mnras]
  {10.1093/mnras/stv1414}, \href
  {http://adsabs.harvard.edu/abs/2015MNRAS.452.2034R} {452, 2034}

\bibitem[\protect\citeauthoryear{{Reddy}, {Steidel}, {Pettini}, {Bogosavljevic}
   \& {Shapley}}{{Reddy} et~al.}{2016}]{2016arXiv160603452R}
{Reddy} N.~A.,  {Steidel} C.~C.,  {Pettini} M.,  {Bogosavljevic} M.,
  {Shapley} A.,  2016, preprint, \href
  {http://adsabs.harvard.edu/abs/2016arXiv160603452R} {} (\mn@eprint {arXiv}
  {1606.03452})

\bibitem[\protect\citeauthoryear{{Robertson} et~al.,}{{Robertson}
  et~al.}{2013}]{2013ApJ...768...71R}
{Robertson} B.~E.,  et~al., 2013, \mn@doi [\apj] {10.1088/0004-637X/768/1/71},
  \href {http://adsabs.harvard.edu/abs/2013ApJ...768...71R} {768, 71}

\bibitem[\protect\citeauthoryear{{Robertson}, {Ellis}, {Furlanetto}  \&
  {Dunlop}}{{Robertson} et~al.}{2015}]{2015ApJ...802L..19R}
{Robertson} B.~E.,  {Ellis} R.~S.,  {Furlanetto} S.~R.,   {Dunlop} J.~S.,
  2015, \mn@doi [\apjl] {10.1088/2041-8205/802/2/L19}, \href
  {http://adsabs.harvard.edu/abs/2015ApJ...802L..19R} {802, L19}

\bibitem[\protect\citeauthoryear{{Rorai}, {Hennawi}  \& {White}}{{Rorai}
  et~al.}{2013}]{2013ApJ...775...81R}
{Rorai} A.,  {Hennawi} J.~F.,   {White} M.,  2013, \mn@doi [\apj]
  {10.1088/0004-637X/775/2/81}, \href
  {http://adsabs.harvard.edu/abs/2013ApJ...775...81R} {775, 81}

\bibitem[\protect\citeauthoryear{{Rosdahl}, {Blaizot}, {Aubert}, {Stranex}  \&
  {Teyssier}}{{Rosdahl} et~al.}{2013}]{2013MNRAS.436.2188R}
{Rosdahl} J.,  {Blaizot} J.,  {Aubert} D.,  {Stranex} T.,   {Teyssier} R.,
  2013, \mn@doi [\mnras] {10.1093/mnras/stt1722}, \href
  {http://adsabs.harvard.edu/abs/2013MNRAS.436.2188R} {436, 2188}

\bibitem[\protect\citeauthoryear{{Shapley}, {Steidel}, {Strom},
  {Bogosavljevi{\'c}}, {Reddy}, {Siana}, {Mostardi}  \& {Rudie}}{{Shapley}
  et~al.}{2016}]{2016ApJ...826L..24S}
{Shapley} A.~E.,  {Steidel} C.~C.,  {Strom} A.~L.,  {Bogosavljevi{\'c}} M.,
  {Reddy} N.~A.,  {Siana} B.,  {Mostardi} R.~E.,   {Rudie} G.~C.,  2016,
  \mn@doi [\apjl] {10.3847/2041-8205/826/2/L24}, \href
  {http://adsabs.harvard.edu/abs/2016ApJ...826L..24S} {826, L24}

\bibitem[\protect\citeauthoryear{{Siana} et~al.,}{{Siana}
  et~al.}{2015}]{2015ApJ...804...17S}
{Siana} B.,  et~al., 2015, \mn@doi [\apj] {10.1088/0004-637X/804/1/17}, \href
  {http://adsabs.harvard.edu/abs/2015ApJ...804...17S} {804, 17}

\bibitem[\protect\citeauthoryear{{Songaila} \& {Cowie}}{{Songaila} \&
  {Cowie}}{2010}]{2010ApJ...721.1448S}
{Songaila} A.,  {Cowie} L.~L.,  2010, \mn@doi [\apj]
  {10.1088/0004-637X/721/2/1448}, \href
  {http://adsabs.harvard.edu/abs/2010ApJ...721.1448S} {721, 1448}

\bibitem[\protect\citeauthoryear{{Springel} et~al.,}{{Springel}
  et~al.}{2005}]{2005Natur.435..629S}
{Springel} V.,  et~al., 2005, \mn@doi [\nat] {10.1038/nature03597}, \href
  {http://adsabs.harvard.edu/abs/2005Natur.435..629S} {435, 629}

\bibitem[\protect\citeauthoryear{{Stark} et~al.,}{{Stark}
  et~al.}{2015}]{2015MNRAS.454.1393S}
{Stark} D.~P.,  et~al., 2015, \mn@doi [\mnras] {10.1093/mnras/stv1907}, \href
  {http://adsabs.harvard.edu/abs/2015MNRAS.454.1393S} {454, 1393}

\bibitem[\protect\citeauthoryear{{Stark} et~al.,}{{Stark}
  et~al.}{2016}]{2016arXiv160601304S}
{Stark} D.~P.,  et~al., 2016, preprint, \href
  {http://adsabs.harvard.edu/abs/2016arXiv160601304S} {} (\mn@eprint {arXiv}
  {1606.01304})

\bibitem[\protect\citeauthoryear{{Telfer}, {Zheng}, {Kriss}  \&
  {Davidsen}}{{Telfer} et~al.}{2002}]{2002ApJ...565..773T}
{Telfer} R.~C.,  {Zheng} W.,  {Kriss} G.~A.,   {Davidsen} A.~F.,  2002, \mn@doi
  [\apj] {10.1086/324689}, \href
  {http://cdsads.u-strasbg.fr/abs/2002ApJ...565..773T} {565, 773}

\bibitem[\protect\citeauthoryear{{Teyssier}}{{Teyssier}}{2002}]{2002A&A...385..337T}
{Teyssier} R.,  2002, \mn@doi [\aap] {10.1051/0004-6361:20011817}, \href
  {http://cdsads.u-strasbg.fr/abs/2002A\%26A...385..337T} {385, 337}

\bibitem[\protect\citeauthoryear{{Theuns}, {Leonard}, {Efstathiou}, {Pearce}
  \& {Thomas}}{{Theuns} et~al.}{1998}]{1998MNRAS.301..478T}
{Theuns} T.,  {Leonard} A.,  {Efstathiou} G.,  {Pearce} F.~R.,   {Thomas}
  P.~A.,  1998, \mn@doi [\mnras] {10.1046/j.1365-8711.1998.02040.x}, \href
  {http://adsabs.harvard.edu/abs/1998MNRAS.301..478T} {301, 478}

\bibitem[\protect\citeauthoryear{{Theuns}, {Mo}  \& {Schaye}}{{Theuns}
  et~al.}{2001}]{2001MNRAS.321..450T}
{Theuns} T.,  {Mo} H.~J.,   {Schaye} J.,  2001, \mn@doi [\mnras]
  {10.1046/j.1365-8711.2001.04026.x}, \href
  {http://adsabs.harvard.edu/abs/2001MNRAS.321..450T} {321, 450}

\bibitem[\protect\citeauthoryear{{Vanden Berk} et~al.,}{{Vanden Berk}
  et~al.}{2001}]{2001AJ....122..549V}
{Vanden Berk} D.~E.,  et~al., 2001, \mn@doi [\aj] {10.1086/321167}, \href
  {http://cdsads.u-strasbg.fr/abs/2001AJ....122..549V} {122, 549}

\bibitem[\protect\citeauthoryear{{Viel}, {Becker}, {Bolton}  \&
  {Haehnelt}}{{Viel} et~al.}{2013}]{2013PhRvD..88d3502V}
{Viel} M.,  {Becker} G.~D.,  {Bolton} J.~S.,   {Haehnelt} M.~G.,  2013, \mn@doi
  [\prd] {10.1103/PhysRevD.88.043502}, \href
  {http://adsabs.harvard.edu/abs/2013PhRvD..88d3502V} {88, 043502}

\bibitem[\protect\citeauthoryear{{Wise}, {Demchenko}, {Halicek}, {Norman},
  {Turk}, {Abel}  \& {Smith}}{{Wise} et~al.}{2014}]{2014MNRAS.442.2560W}
{Wise} J.~H.,  {Demchenko} V.~G.,  {Halicek} M.~T.,  {Norman} M.~L.,  {Turk}
  M.~J.,  {Abel} T.,   {Smith} B.~D.,  2014, \mn@doi [\mnras]
  {10.1093/mnras/stu979}, \href
  {http://adsabs.harvard.edu/abs/2014MNRAS.442.2560W} {442, 2560}

\bibitem[\protect\citeauthoryear{{Worseck} et~al.,}{{Worseck}
  et~al.}{2014}]{2014MNRAS.445.1745W}
{Worseck} G.,  et~al., 2014, \mn@doi [\mnras] {10.1093/mnras/stu1827}, \href
  {http://adsabs.harvard.edu/abs/2014MNRAS.445.1745W} {445, 1745}

\bibitem[\protect\citeauthoryear{{Wyithe} \& {Bolton}}{{Wyithe} \&
  {Bolton}}{2011}]{2011MNRAS.412.1926W}
{Wyithe} J.~S.~B.,  {Bolton} J.~S.,  2011, \mn@doi [\mnras]
  {10.1111/j.1365-2966.2010.18030.x}, \href
  {http://adsabs.harvard.edu/abs/2011MNRAS.412.1926W} {412, 1926}

\bibitem[\protect\citeauthoryear{{Wyithe}, {Bolton}  \& {Haehnelt}}{{Wyithe}
  et~al.}{2008}]{2008MNRAS.383..691W}
{Wyithe} J.~S.~B.,  {Bolton} J.~S.,   {Haehnelt} M.~G.,  2008, \mn@doi [\mnras]
  {10.1111/j.1365-2966.2007.12578.x}, \href
  {http://adsabs.harvard.edu/abs/2008MNRAS.383..691W} {383, 691}

\bibitem[\protect\citeauthoryear{{Xu}, {Wise}, {Norman}, {Ahn}  \&
  {O'Shea}}{{Xu} et~al.}{2016}]{2016arXiv160407842X}
{Xu} H.,  {Wise} J.~H.,  {Norman} M.~L.,  {Ahn} K.,   {O'Shea} B.~W.,  2016,
  preprint, \href {http://adsabs.harvard.edu/abs/2016arXiv160407842X} {}
  (\mn@eprint {arXiv} {1604.07842})

\bibitem[\protect\citeauthoryear{{Yoshiura}, {Hasegawa}, {Ichiki}, {Tashiro},
  {Shimabukuro}  \& {Takahashi}}{{Yoshiura} et~al.}{2016}]{2016arXiv160204407Y}
{Yoshiura} S.,  {Hasegawa} K.,  {Ichiki} K.,  {Tashiro} H.,  {Shimabukuro} H.,
   {Takahashi} K.,  2016, preprint, \href
  {http://adsabs.harvard.edu/abs/2016arXiv160204407Y} {} (\mn@eprint {arXiv}
  {1602.04407})

\makeatother
\end{thebibliography}

\appendix

\section{The varying mean free path}
\label{lambda_mfp_delta}

In section \ref{mfpsubsection} we had discussed our modelling of the mean-free path.  
Here, we show in more detail how the  photoionization rate as well as the mean free path 
vary spatially in our  different models. The case of a constant mean free path is compared 
to our models where the  mean free path depends on  the local value of 
$\mathrm{\Gamma_{12}}$ as  a power law (see \citealt{2016MNRAS.460.1328D}),

\begin{equation}
 \mathrm{\lambda_{mfp}(\Gamma)=\lambda_0(\Gamma/\Gamma_0)^{2/3}\Delta^{-\gamma}}.
\label{furlanet_gamma2}
\end{equation}

We have considered  two different values of $\mathrm{\gamma}$ to test the dependence of the mean free path on local overdensity.
First we tested $\mathrm{\gamma}=1$ as in \citealt{2016MNRAS.460.1328D} and  then 
we also investigated models with  $\mathrm{\gamma}=0.4$ which is the value we found in our simulations.

We note here that we used the dark matter overdensity from the Millennium simulation to estimate  $\mathrm{\Delta}$ and that 
we have smoothed the density field with a  20 cMpc and  5 cMpc wide top-hat filter.
\citealt{2016MNRAS.460.1328D} have argued that  such a  coarse resolution is sufficient (they have used 5 cMpc) as  it resolves the typical mean 
free path at the considered redshifts. We find very little differences by smoothing the field on  scales of 5 or 20 cMpc.

\begin{figure*}
   \begin{center}
      \includegraphics[width=17cm,height=22cm,keepaspectratio]{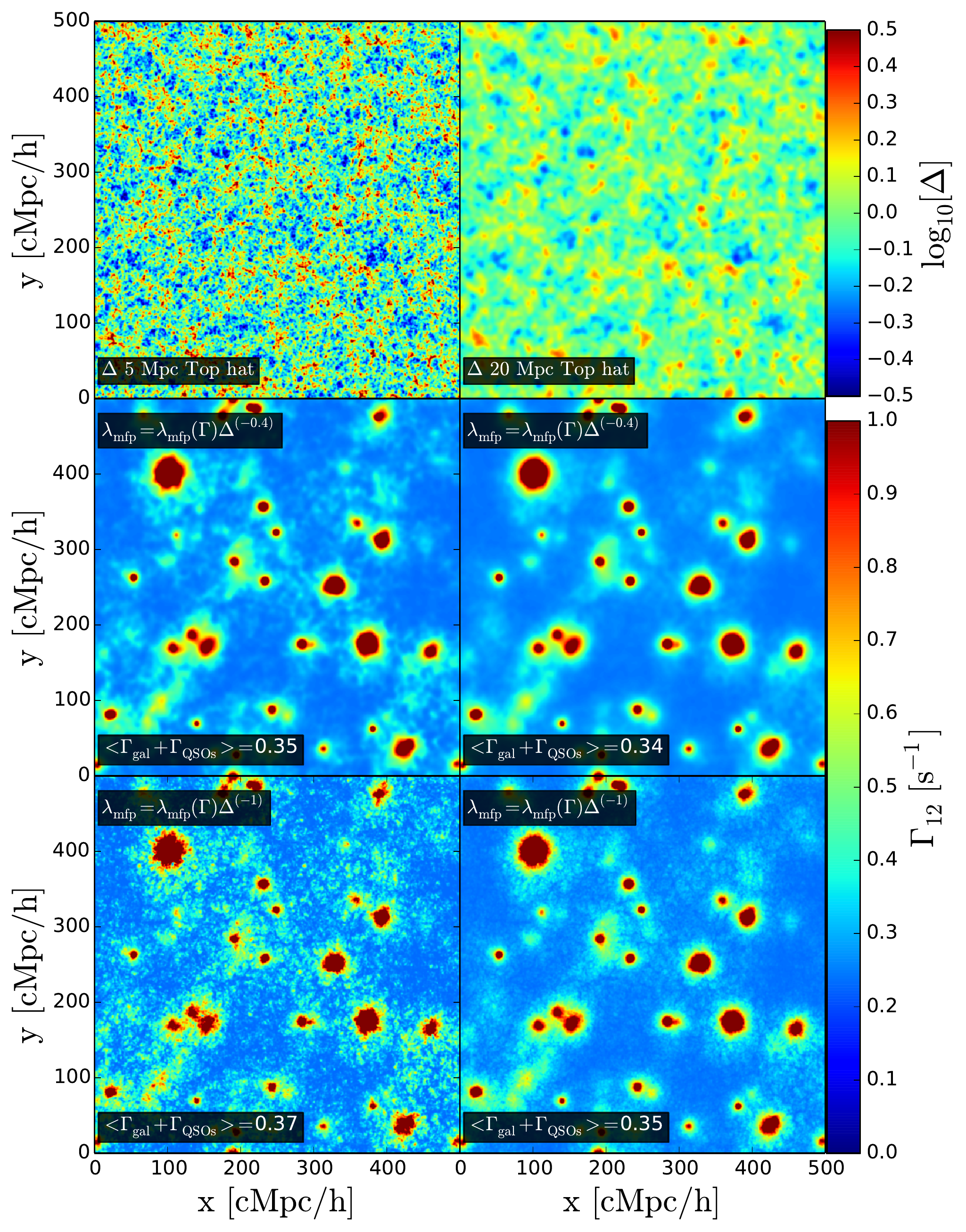}    
  \caption{The spatial distribution of the photoionization rate $\mathrm{\Gamma_{12}}$ in a slice of 976.5625 cMpc/h in our different models 
with  the luminosity function of \citealt{2015AA...578A..83G}. The left column shows the case for the density field $\mathrm{\Delta}$ 
smoothed with a  5cMpc wide (top left panel) top hat filter and  20 cMpc  wide top hat filter (top right panel).
The second row shows $\mathrm{\Gamma}$ for an overdensity dependence of the mean free path  $\propto \mathrm{\Delta^{-0.4}}$ while the bottom row is for $\mathrm{\Delta^{-1}}$. 
}
    \label{gamma_diff_model_map}
  \end{center}
 \end{figure*}

Fig. \ref{gamma_diff_model_map} shows the spatial distribution of $\mathrm{\Gamma_{12}}$ for  our four different models tested.
For illustrative purposes we also show the overdensity field smoothed with a  5 or 20 cMpc wide filter in the two upper panels.
Overall, the power law dependence with overdensity of the mean free path as well as the smoothing adopted for the overdensity field only
have a small effect on  the global photoionization-rate field. A higher granularity is found when smoothing  with the 5 cMpc wide filter  
compared to  the 20 cMpc wide filter, but similar values of $\left<\Gamma\right>$ 
are found whatever the assumed power law index for the overdensity dependence of the mean free path.
Raising  the power law index from 0.4 to 1 changes the average value of $\left<\Gamma\right>$ for both smoothing scales
only very moderately.

\begin{figure*}
   \begin{center}
      \includegraphics[width=\textwidth,height=\textheight,keepaspectratio]{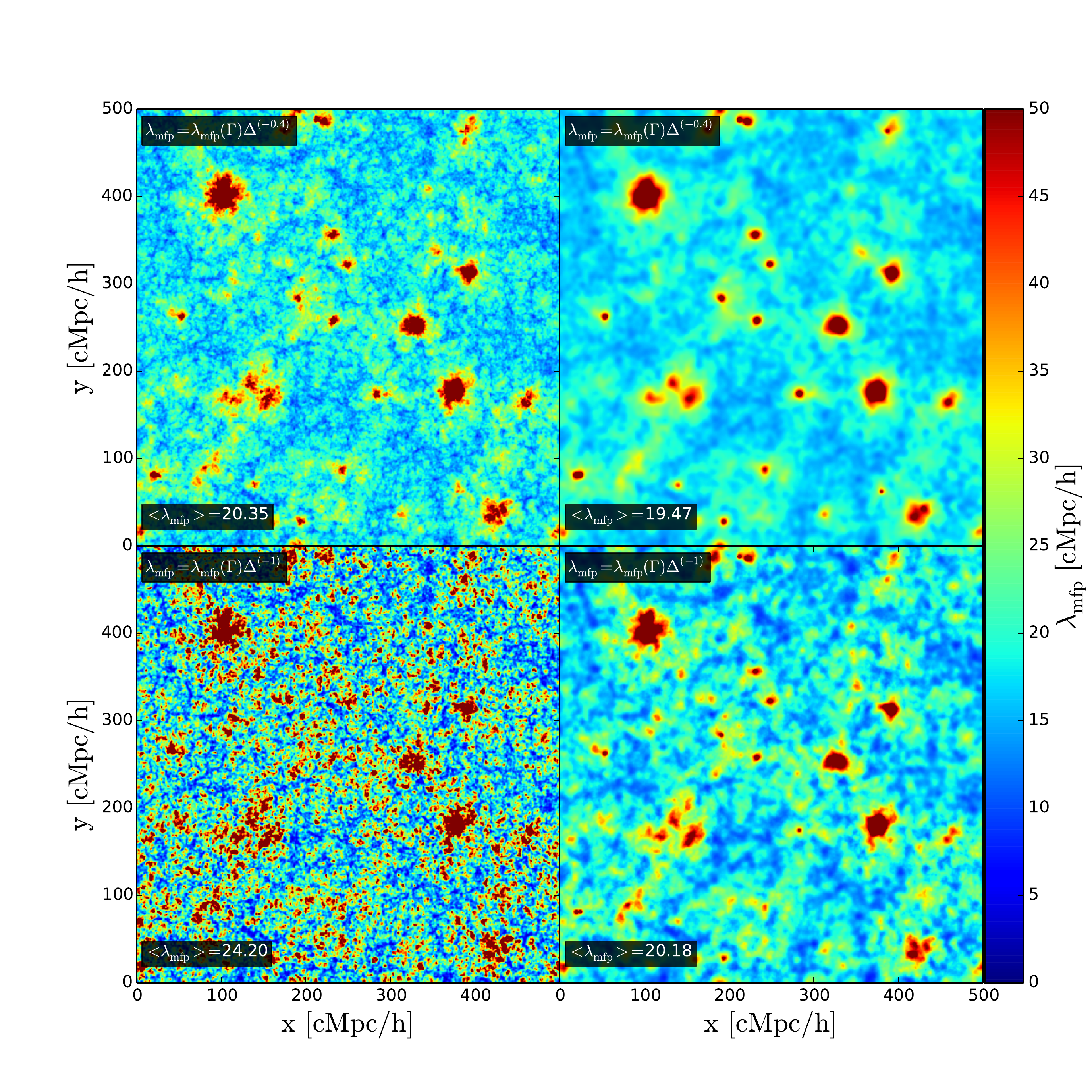}    
  \caption{The spatial distribution of the mean free path $\mathrm{\lambda_{mfp}}$  for  our different models with 
the luminosity function of \citet{2015AA...578A..83G}.
The left column shows the case in which the density field $\mathrm{\Delta}$ has been smoothed 
on a 5 cMpc scale while the right column is based on the 20 cMpc smoothing.
The top row shows $\mathrm{\Gamma}$ for an overdensity dependence of the mean free path  
$\propto \mathrm{\Delta^{-0.4}}$ while the bottom row is for $\mathrm{\Delta^{-1}}$.}
    \label{mfp_diff_model_map}
  \end{center}
 \end{figure*}

The same is true for the  spatial distribution of the mean free path itself shown in Fig. \ref{mfp_diff_model_map}.
A smaller value of the smoothing scale of 5 cMpc and a steeper dependence of overdensity with $\gamma =1$ lead to a slightly larger average 
value of the mean free path compared to the other cases. 
But this small difference in the global mean free path translates only to rather small changes in the photoionization-rate maps, which in turn only moderately change the $\mathrm{\tau_{eff}}$ PDF as seen in Sect. \ref{PDFteff}.

\end{document}